\documentclass[letterpaper,twocolumn,10pt]{article}
\pdfoutput=1
\usepackage{usenix}
\microtypecontext{spacing=nonfrench}
\usepackage{tikz}
\usepackage{amsmath}
\usepackage{amssymb}
\usepackage{algorithm}
\usepackage{algpseudocode} 
\makeatletter
\def\fps@algorithm{t}
\makeatother
\usepackage{filecontents}
\usepackage{booktabs}
\usepackage{multirow}
\usepackage{array}
\usepackage{tabularx}
\usepackage{appendix}
\usepackage{siunitx} 
\usepackage{subfig}
\usepackage{url}
\usepackage[table]{xcolor}
\usepackage{placeins}
\begin{document}

\date{}

\title{\Large \bf BiRD: A Bidirectional Ranking Defense Mechanism for \\ Retrieval Augmented Generation
}

\author{
Chengcai Gao$^{1}$, Zhihong Sun$^{2}$, Xiaochuan Shi$^{1}$, Qiufeng Wang$^{3}$, Chao Liang$^{1}$\\
\small 1 Wuhan University, 2 Naval University of Engineering, 3 Xi'an Jiaotong-Liverpool University\\
\small 1\{2023302051075, shixiaochuan, cliang\}@whu.edu.cn, 2 zhihong.sun@whu.edu.cn, 3 Qiufeng.Wang@xjtlu.edu.cn
}

\maketitle

\begin{abstract}
The growing adoption of Retrieval-Augmented Generation (RAG) has led to a rise in adversarial attacks. Existing defenses, relying on semantic analysis or voting, face a trade-off between high computational cost and limited robustness under strong poisoning attacks. Their fundamental limitation is the exclusive focus on semantic content relevance, while neglecting the retrieval context that is critically defined by ranking structures. To this end, we investigate the bidirectional ranking behavior of poisoned and benign documents, and discover a key discriminative pattern: poisoned documents exhibit significantly stronger alignment between their backward rankings and the query's forward ranking. Capitalizing on this, we propose BiRD, a bidirectional ranking defense mechanism built upon a dual-signal framework that leverages forward ranking to assess semantic content relevance and backward ranking to quantify ranking context consistency. This design directly addresses the fundamental limitation of prior approaches, enabling simultaneous efficiency and robustness. Extensive evaluation across 3 datasets with 3 retrievers and 3 LLMs under 2 attack scenarios validates BiRD's effectiveness. Notably, BiRD reduces the attack success rate of PoisonedRAG by up to 54\% while simultaneously improving task accuracy by up to 56\%, with average additional latency under 1 second.
\end{abstract}
\section{Introduction}
\begin{figure}[ht]
    \centering
    \includegraphics[width=\columnwidth]{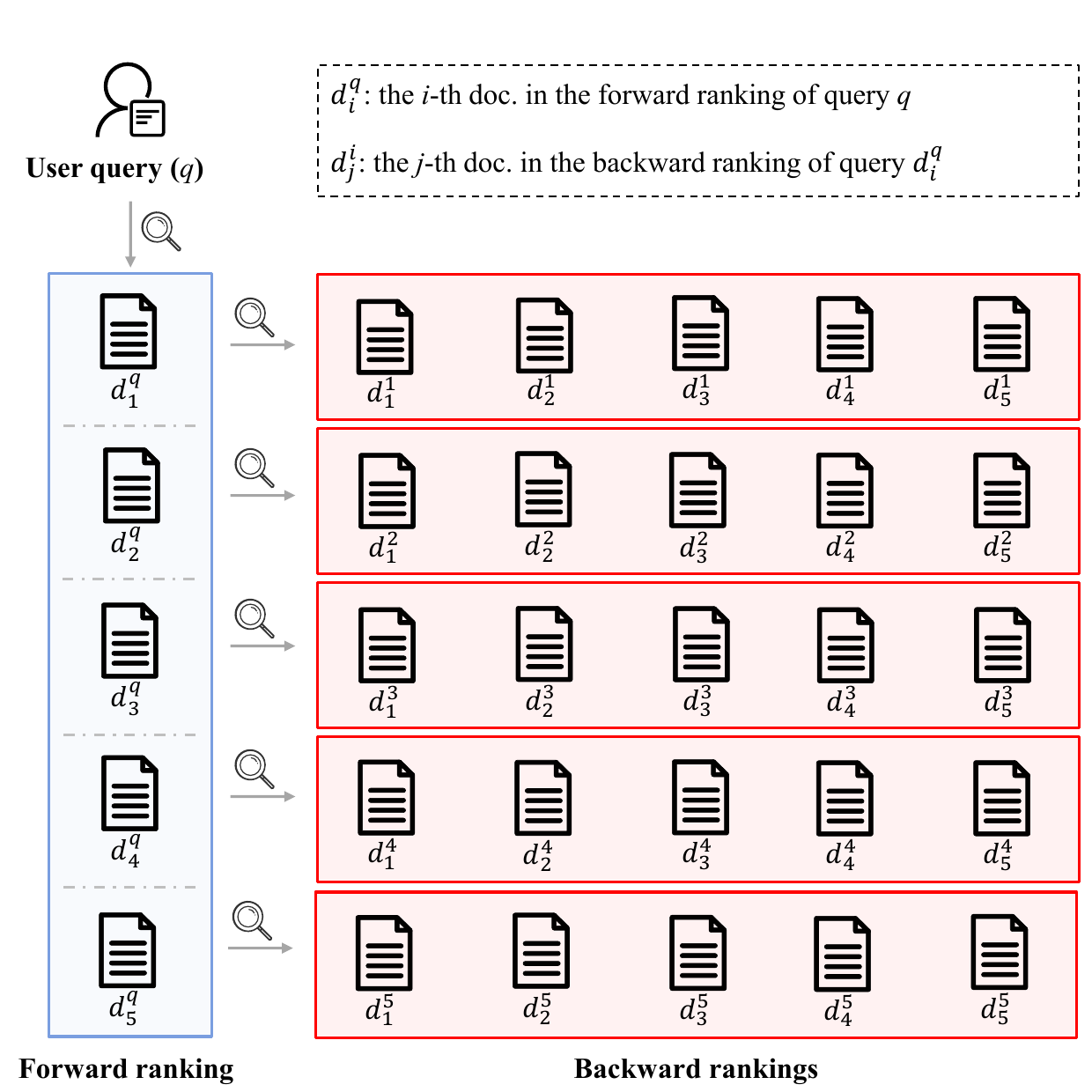}
    \caption{Bidirectional Ranking Defense (BiRD) framework.}
    \label{fig:Comparison_with_previous_work}
\end{figure}
While Large Language Models (LLMs) have significantly advanced natural language processing, they struggle with generating hallucinations \cite{huangSurveyHallucinationLarge2025}  and accessing real-time knowledge \cite{yaoSurveyLargeLanguage2024}. To address these limitations, 
Retrieval-Augmented Generation (RAG) emerges as a pivotal preliminary stage,enhancing the accuracy and reliability of generated content by retrieving relevant information from external knowledge bases as context \cite{gaoRetrievalAugmentedGenerationLarge2024}. However, RAG systems introduce new security vulnerabilities, most notably corpus poisoning attacks \cite{zouPoisonedRAGKnowledgeCorruption2024, zhongPoisoningRetrievalCorpora2023, ben-tovGASLITEingRetrievalExploring2025, songSilentSaboteurImperceptible2025}. Attackers inject a few meticulously crafted malicious documents into the corpus to manipulate retrieval results, thereby inducing downstream LLMs to generate misleading \cite{gongTopicFlipRAGTopicOrientatedAdversarial2025}, biased \cite{bagweYourRAGUnfair2025}, or even harmful content \cite{zouPoisonedRAGKnowledgeCorruption2024}. Therefore, designing robust defense mechanisms to secure the retrieval stage in RAG systems has become a crucial and urgent priority.

Existing defenses against corpus poisoning in RAG systems typically adopt one of three strategies. First, clustering-filtering methods \cite{siSeConRAGTwoStageSemantic2025,zhouTrustRAGEnhancingRobustness2025} identify anomalies by analyzing semantic content, such as entity-intent relationships, but this detailed parsing introduces substantial computational overhead. Building on semantic analysis yet aiming for simpler detection, voting-consistency approaches \cite{shenReliabilityRAGEffectiveProvably2025,xiangCertifiablyRobustRAG2024} rely on cross-verifying outputs across multiple paths. However, their dependence on consensus leaves them vulnerable to high-intensity poisoning attacks that can manipulate the voting outcome. Alternatively, adversarial defense techniques \cite{xianVulnerabilityApplyingRetrievalAugmented2025,wangAstuteRAGOvercoming2025,weiInstructRAGInstructingRetrievalAugmented2024} leverage the internal knowledge of LLMs to filter semantically inconsistent documents. This strategy, however, is inherently limited by the LLM's own knowledge boundaries and susceptibility to deception. Despite these different designs, a fundamental and common shortcoming persists: these methods focus predominantly on analyzing the semantic content within documents, while critically overlooking the retrieval context, a dimension defined by the ranking structures that govern document order during retrieval (Figure \ref{fig:Comparison_with_previous_work}).

\begin{figure}[ht]
\centering
\includegraphics[width=\columnwidth]{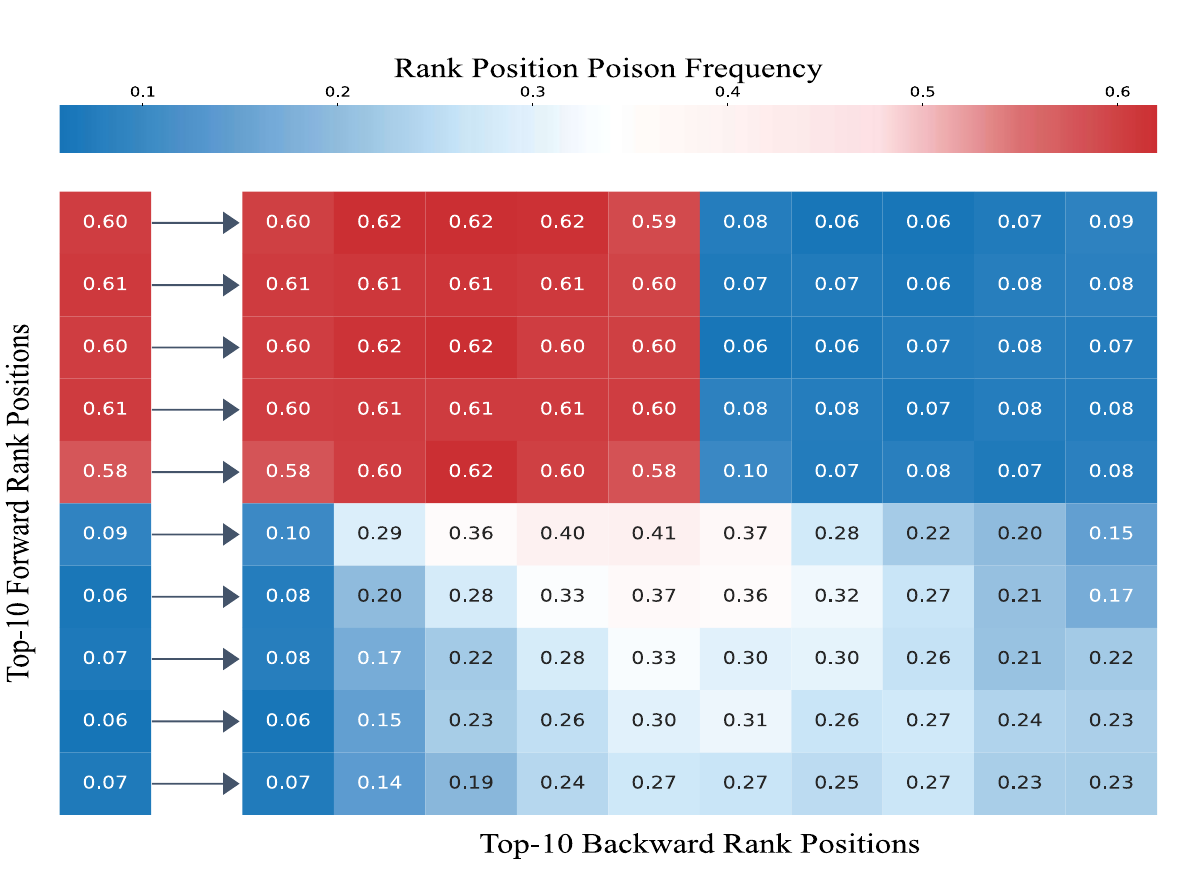}
\caption{Rank Position Poison Frequency Statistical Heatmaps.}
\label{fig:Preliminary_experiment_heatmap}
\end{figure}

To demonstrate its importance, we statistically analyzed the distribution of poisoned documents, specifically their occurrence probability at each ranking position, in bidirectional retrieval experiments conducted across 3 datasets, 3 retrievers, and 2 attack methods (see Section \ref{subsec:pilot-experiments} for details). The results, shown in Figure \ref{fig:Preliminary_experiment_heatmap}, reveal two systematic patterns. First, the left column illustrates how frequently poisoned documents occupy each position in forward retrieval. Second, the right matrix depicts, for each document position treated as a new query in backward retrieval, how likely poisoned documents are to appear at various ranks in the resulting backward list. This observed strong alignment, where poisoned documents not only achieve high forward ranks but also maintain consistent backward-forward ranking relationships, provides a novel, ranking-based signal for defense. This finding motivates the design of a method that jointly evaluates intra-document content relevance and inter-document context consistency.

Building on this insight, we propose BiRD, a bidirectional ranking defense mechanism for RAG. BiRD operates through three stages: (1) the forward retrieval stage assesses semantic content relevance; (2) the backward retrieval stage quantifies ranking context consistency; and (3) the threat filtering stage synthesizes both signals to filter poisoned documents from the top ranked results. By directly leveraging the ranking structures overlooked by prior defenses, BiRD avoids both their high computational costs and vulnerability to strong poisoning attacks, offering an efficient and robust defense for RAG.

The main contributions of this paper can be summarized as follows:
\begin{itemize} \item We discover, for the first time, a key discriminative pattern in bidirectional retrieval: \textit{poisoned documents exhibit significantly stronger alignment between their backward rankings and the query's forward ranking compared to benign documents}. \item We propose BiRD, a novel defense mechanism that introduces and utilizes context consistency, which is measured through bidirectional ranking alignment, in conjunction with content relevance to protect RAG systems against diverse poisoning attacks. \item We conduct extensive experiments across 3 diverse datasets, utilizing 3 different retrievers and 3 LLMs, and under 2 distinct RAG attack scenarios, to comprehensively validate both the effectiveness and robustness of BiRD.\end{itemize}

\section{Related Work}
This section surveys the related work in RAG security research. We begin by introducing the foundational framework of RAG in Section \ref{subsec:RAG}, followed by a review of the relevant work on RAG attacks in Section \ref{subsec:RAG attack} and RAG defenses in Section \ref{subsec:RAG defenses}, respectively. Finally, we discuss the broad applications of bidirectional search in Section \ref{subsec:Bidirectional Search}. 
\subsection{RAG System}
\label{subsec:RAG}
The standard RAG workflow primarily consists of two stages: retrieval and generation. During the retrieval stage, the system filters a set of top-ranked documents from an external corpus based on semantic similarity (e.g., cosine or dot product) between the user's query and documents. In the generation stage, the large language model synthesizes a final answer by using both the original query and the retrieved document set as context\cite{yaoSurveyLargeLanguage2024,maQueryRewritingRetrievalAugmented2023}.

By leveraging external knowledge, RAG enhances the accuracy and breadth of generated content\cite{lewisRetrievalAugmentedGenerationKnowledgeIntensive2020,zhangSirensSongAI2025}. Its key advantage is the direct use of up-to-date or specialized information absent from training data, which helps reduce LLM hallucinations. The system also offers scalability and flexibility\cite{xuSimRAGSelfImprovingRetrievalAugmented2025}, as knowledge can be updated by refreshing the retrieval corpus without retraining the generative model.

However, beyond its performance benefits, RAG also introduces new attack surfaces. The vulnerability of RAG mainly manifests in the fragility of the retriever during the retrieval phase. When a query is maliciously tampered with or the corpus is poisoned, the document ranking produced by the retriever may include irrelevant or malicious documents\cite{xianVulnerabilityApplyingRetrievalAugmented2025}.

\subsection{RAG Attack}
\label{subsec:RAG attack}
The core objective of RAG attacks is to manipulate the system's input or knowledge base in order to mislead its output. At the input level, attackers may attempt to embed imperceptible instructions within user queries or retrieved content to hijack LLM behavior\cite{zhangHijackRAGHijackingAttacks2024,jiaoPRAttackCoordinatedPromptRAG2025}. In practice, however, directly accessing or manipulating user queries is often difficult.

Considering the static nature of the knowledge base and the dynamic nature of queries in RAG systems, it is easier for attackers to perform corpus poisoning attacks. Among these, corpus poisoning is the most prevalent and poses a particularly realistic threat to RAG security. Attackers represented by PoisonedRAG\cite{zouPoisonedRAGKnowledgeCorruption2024} directly inject maliciously crafted documents into the knowledge base, either through adversarial optimization to boost the retrieval ranking of irrelevant or harmful documents, or by embedding specific biases\cite{bagweYourRAGUnfair2025} or stances\cite{gongTopicFlipRAGTopicOrientatedAdversarial2025} within documents to systematically distort model outputs. GASLITE\cite{ben-tovGASLITEingRetrievalExploring2025} proposes white-box attacks on RAG to evaluate its vulnerabilities under worst-case scenarios. ReGENT\cite{songSilentSaboteurImperceptible2025} is a reinforcement learning-based framework that monitors the interaction dynamics between the attacker and the target RAG system, continuously refining poisoned documents through relevance, generation, and naturalness rewards. Given the prevalence of corpus poisoning attacks, this paper will primarily evaluate the effectiveness of existing defense measures against corpus poisoning attacks.

\subsection{RAG Defense}
\label{subsec:RAG defenses}
\begin{figure*}[!t]
    \centering
    \includegraphics[width=0.8\textwidth]{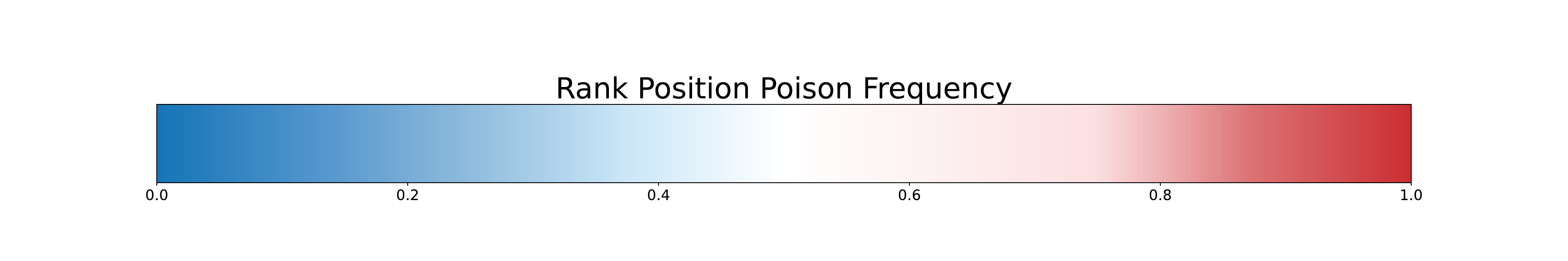} \\
    \vspace{2mm}

    \subfloat[HotpotQA-ANCE]{
        \includegraphics[width=0.31\textwidth]{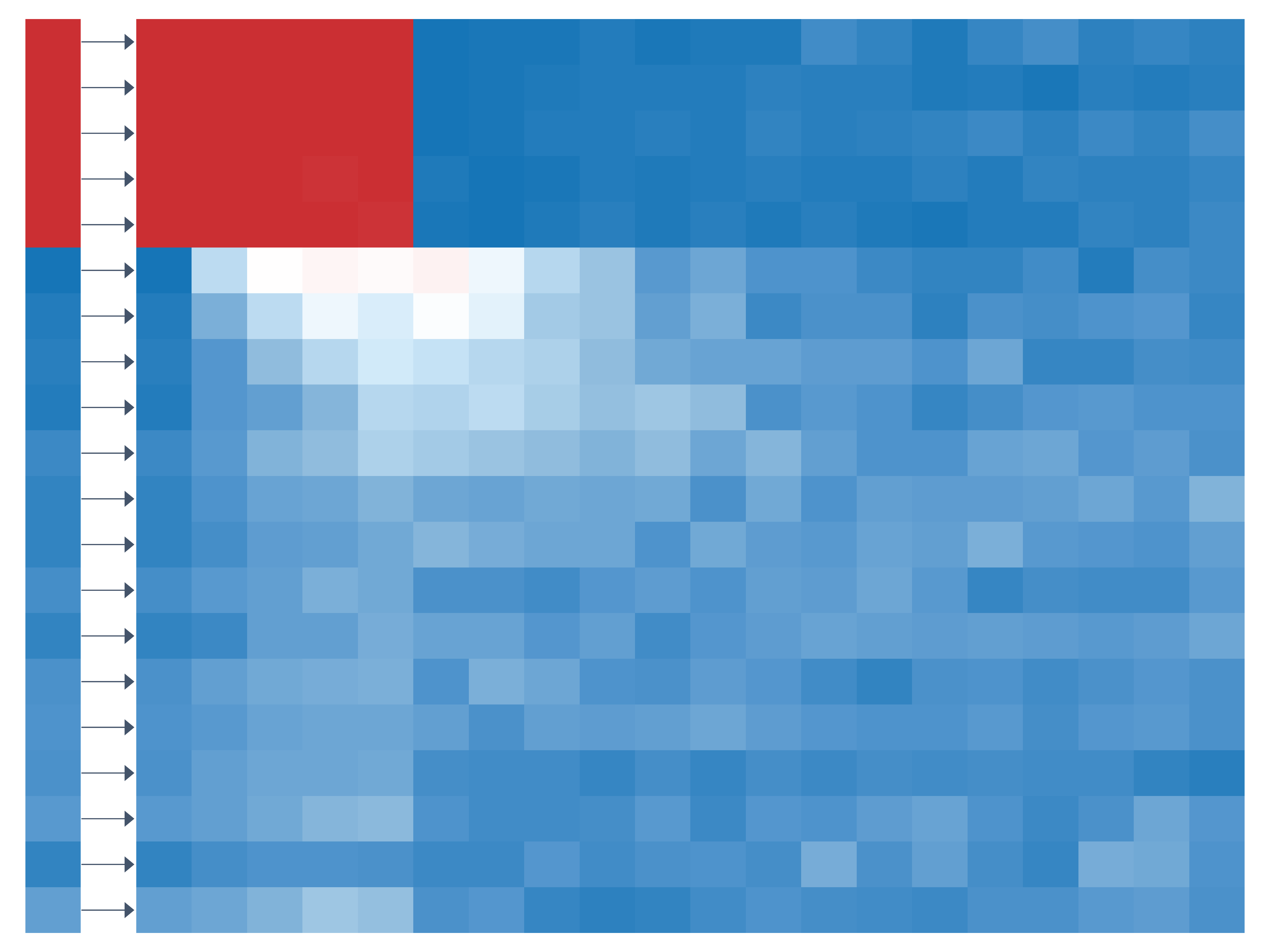}
    } \hfill
    \subfloat[HotpotQA-Contriever]{
        \includegraphics[width=0.31\textwidth]{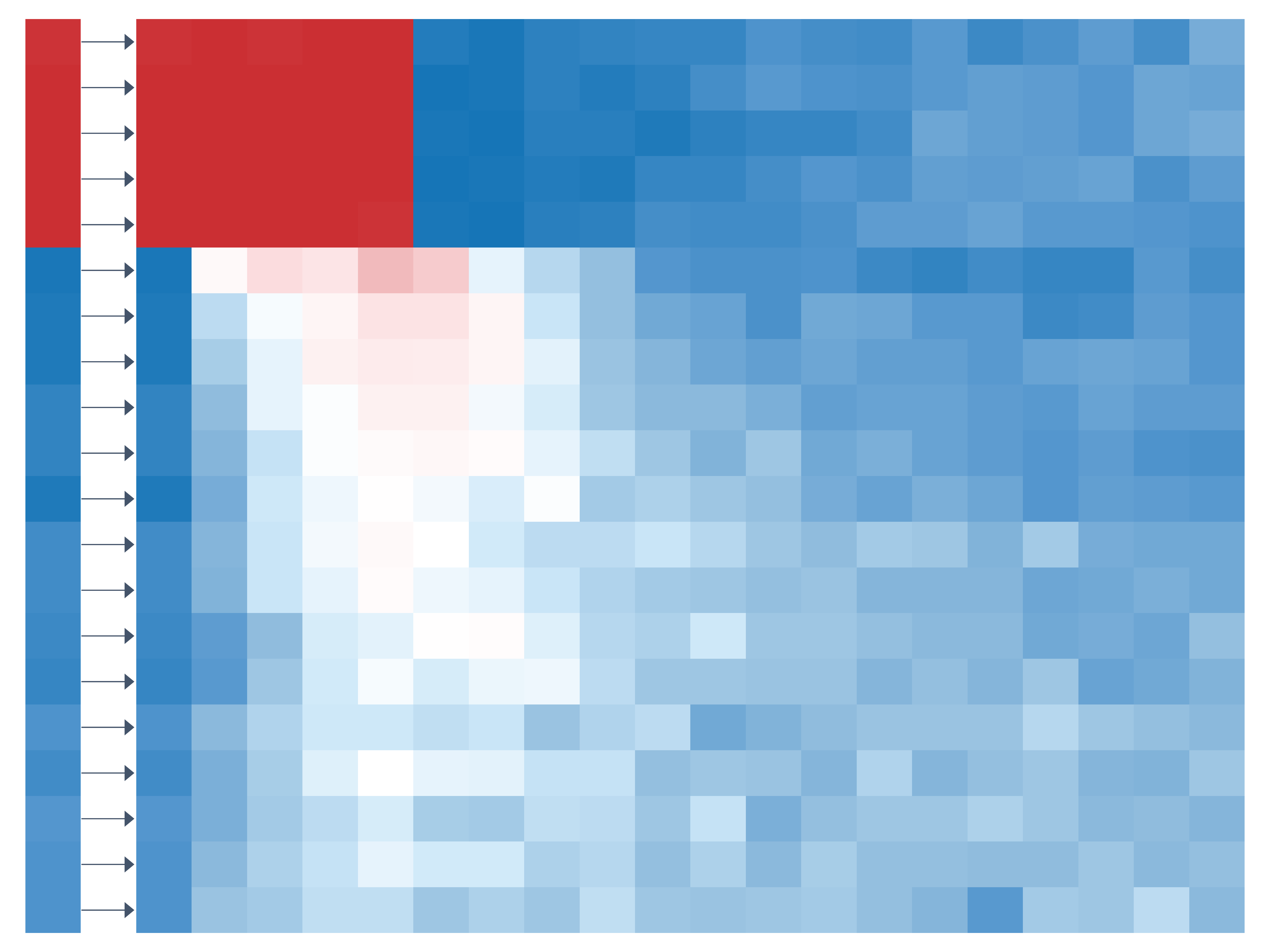}
    } \hfill
    \subfloat[HotpotQA-DPR]{
        \includegraphics[width=0.31\textwidth]{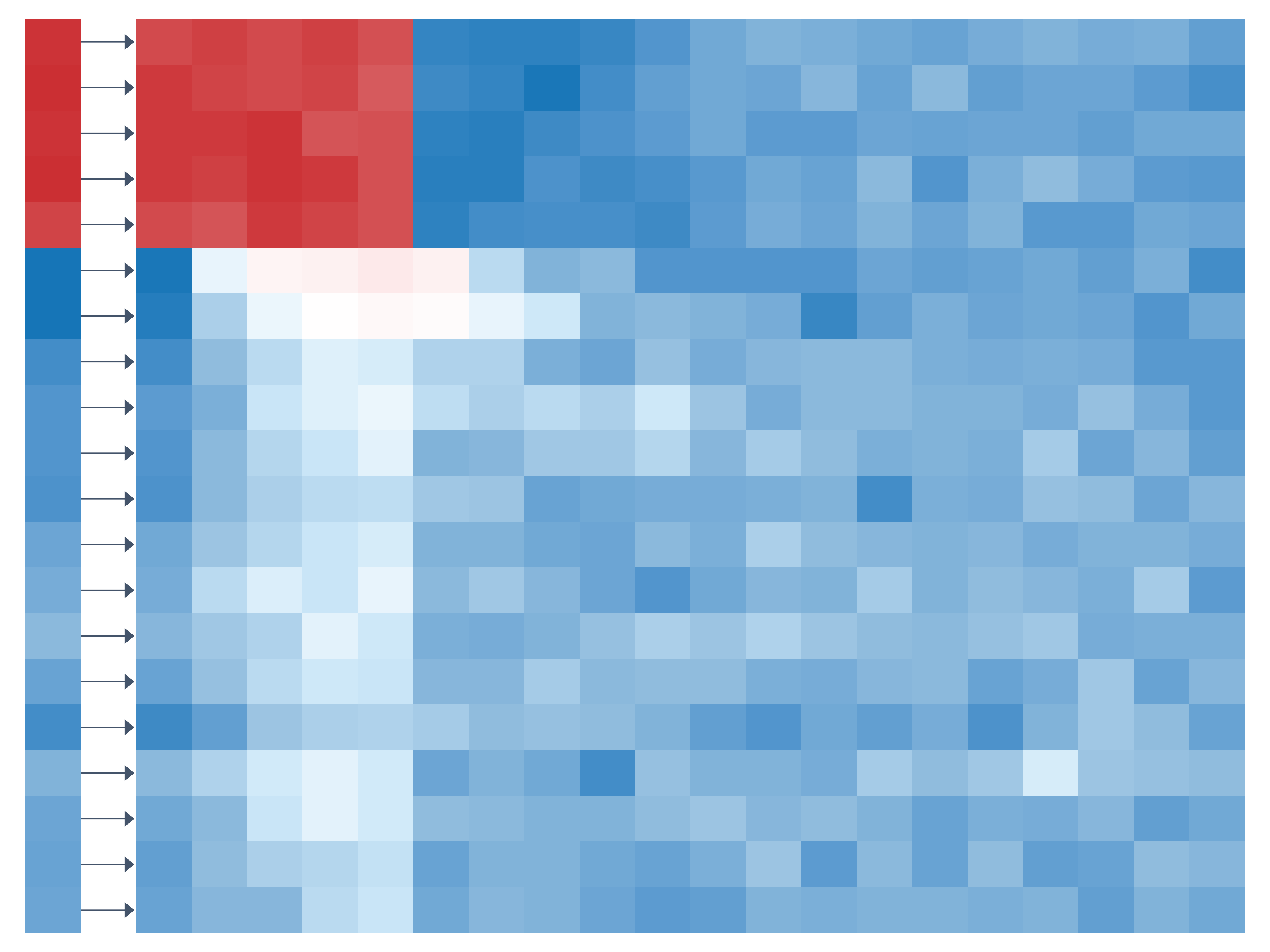}
    }

    \vspace{2mm} 

    \subfloat[HotpotQA-Contriever]{
        \includegraphics[width=0.31\textwidth]{./figs/hotpotqa_contriever_results_heatmap.pdf}
    } \hfill
    \subfloat[MSMARCO-Contriever]{
        \includegraphics[width=0.31\textwidth]{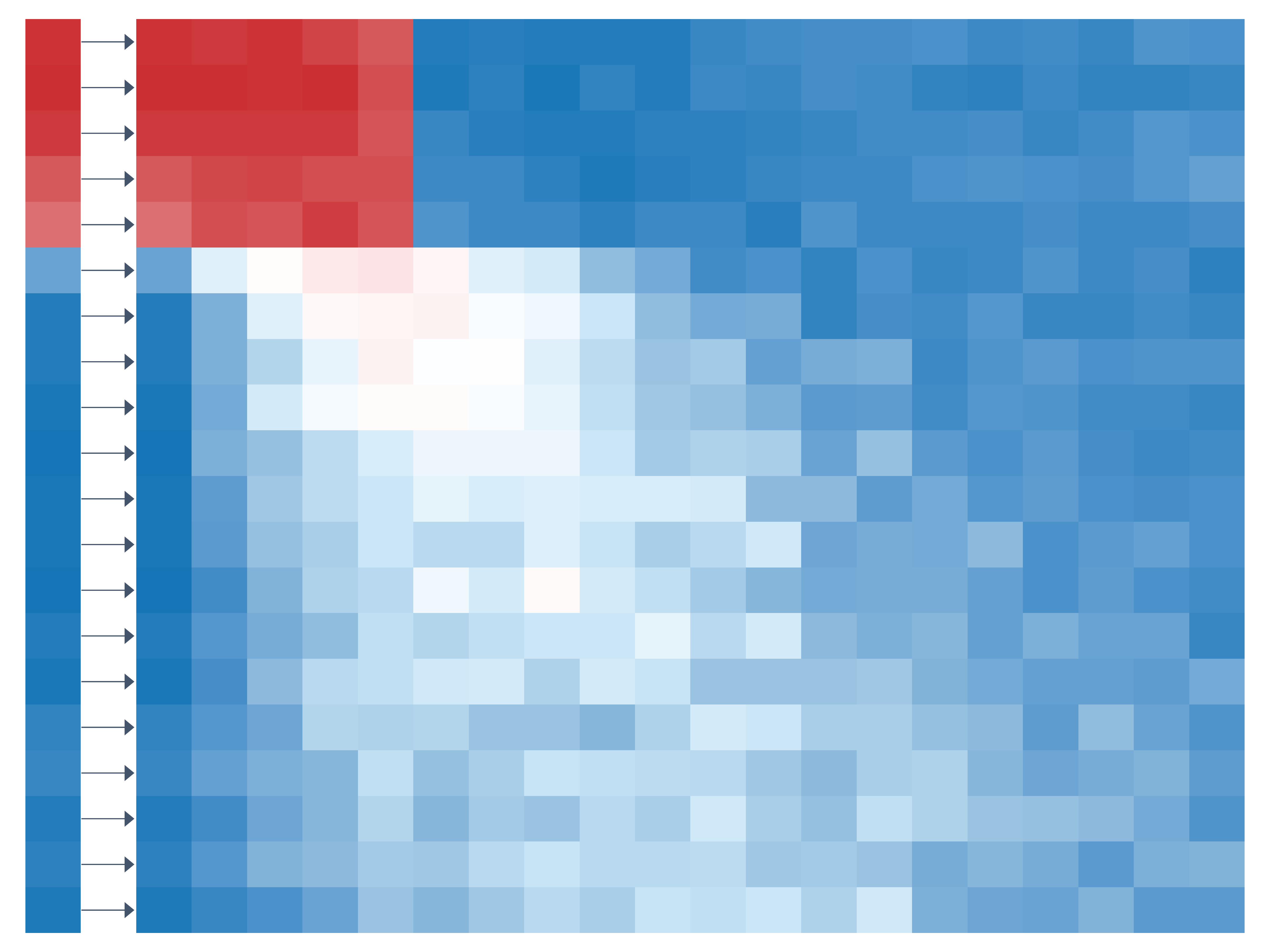}
    } \hfill
    \subfloat[NQ-Contriever]{
        \includegraphics[width=0.31\textwidth]{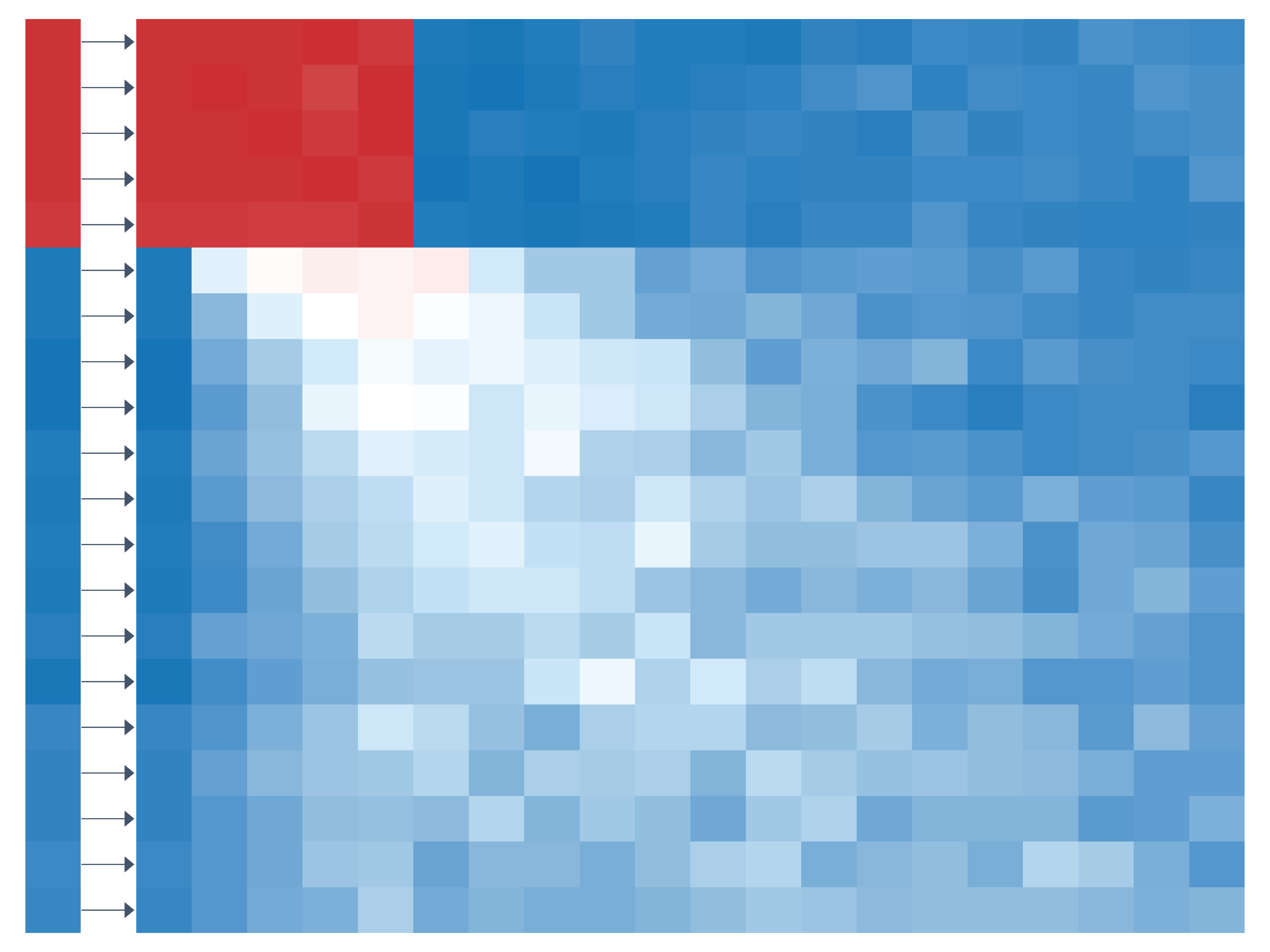}
    }

    \vspace{2mm}

    \subfloat[PoisonedRAG]{
        \includegraphics[width=0.31\textwidth]{./figs/hotpotqa_contriever_results_heatmap.pdf}
    } \hfill
    \subfloat[Topic-Flip-PRO]{
        \includegraphics[width=0.31\textwidth]{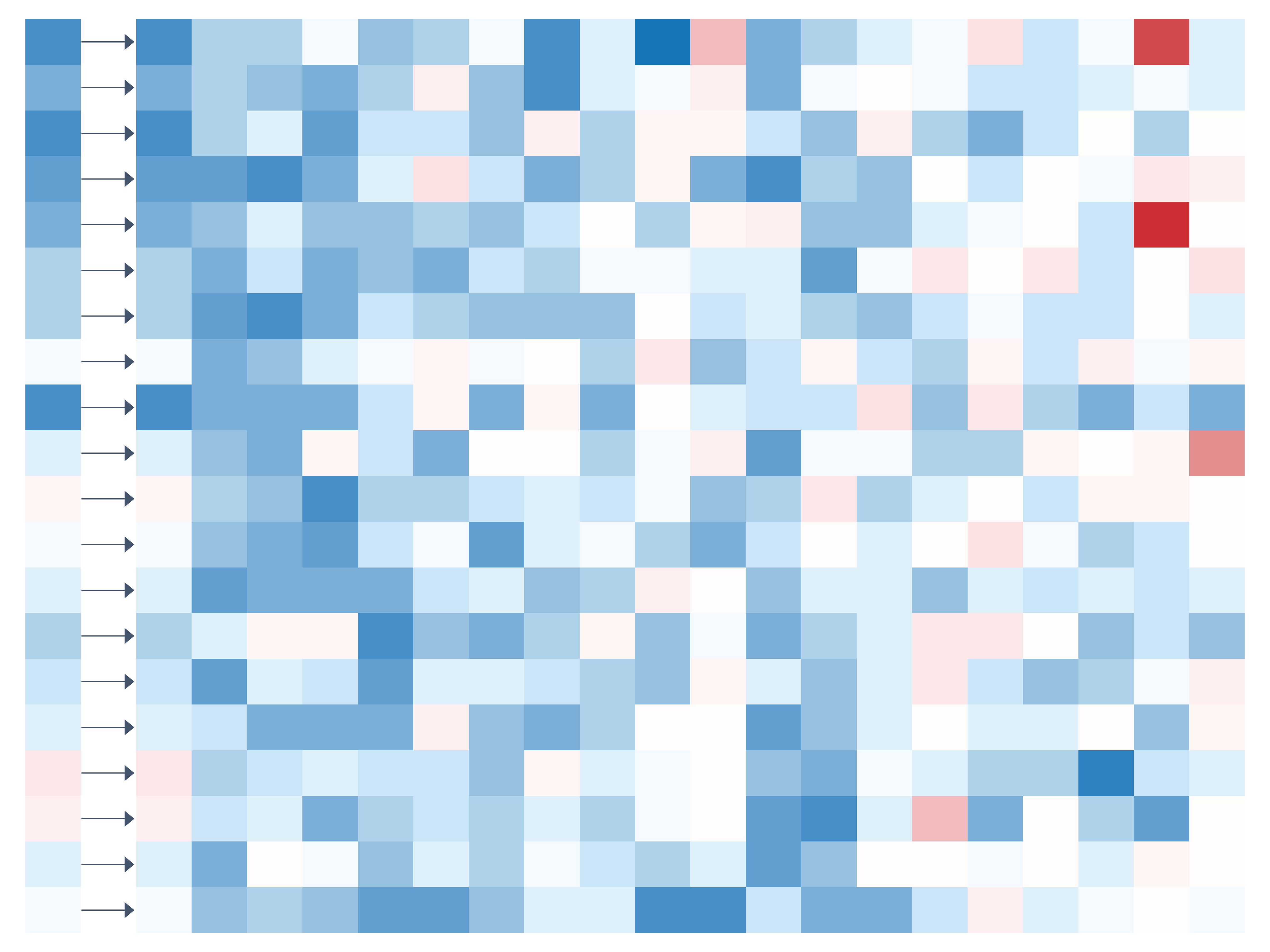}
    } \hfill
    \subfloat[Topic-Flip-CON]{
        \includegraphics[width=0.31\textwidth]{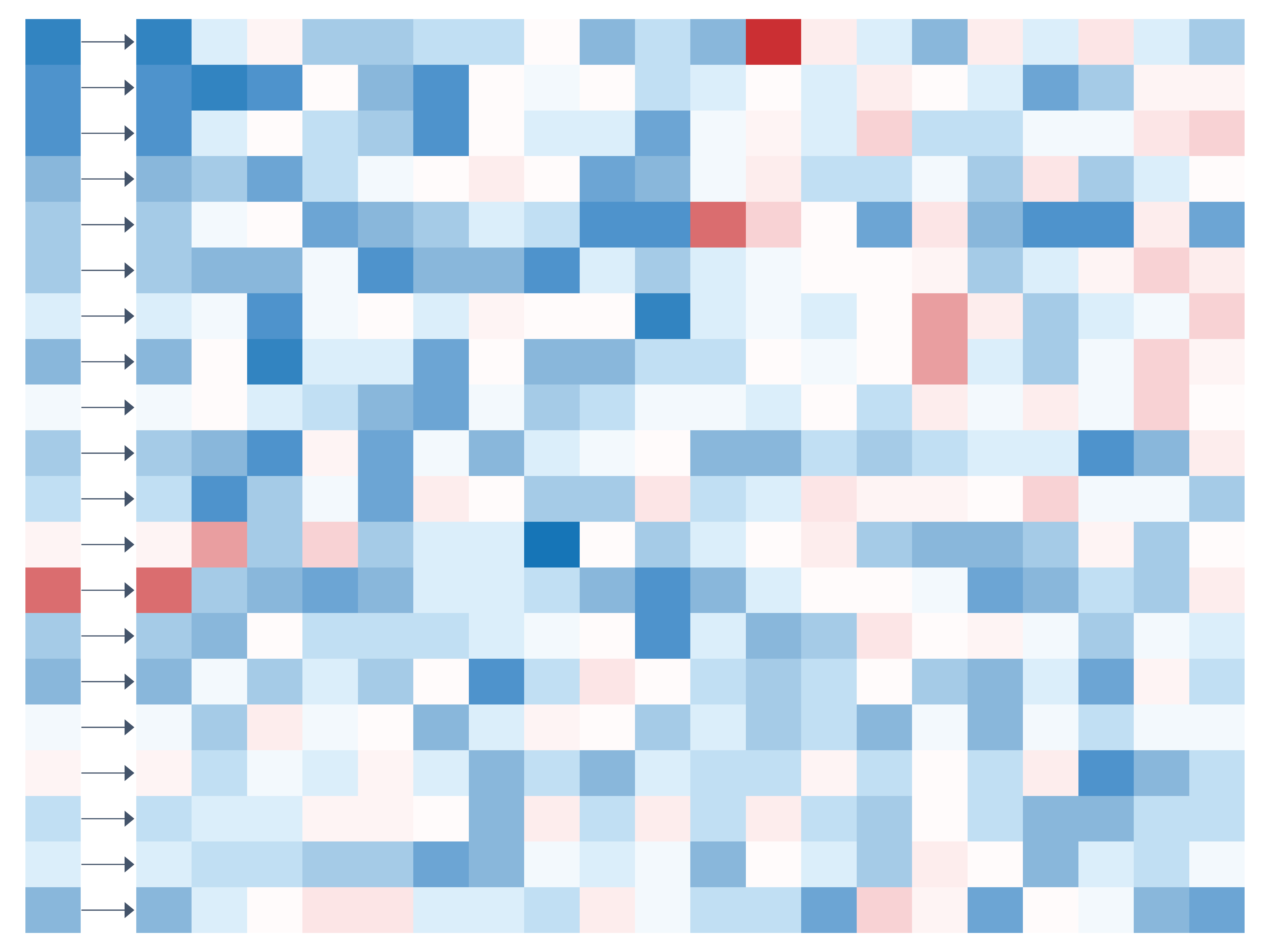}
    }

    \caption{Rank position poison frequency comparison: (Top) different retrievers on HotpotQA; (Middle) Contriever across datasets; (Bottom) comparison of attack algorithms.}
    \label{fig:combined_results}
\end{figure*}
Current defense research primarily develops along three trajectories: some approaches isolate malicious documents through clustering and filtering techniques. For instance, Se-ConRAG \cite{siSeConRAGTwoStageSemantic2025} relies on an external LLMs to construct and compare semantic graphs, while TrustRAG \cite{zhouTrustRAGEnhancingRobustness2025} utilizes an internal LLM for conflict resolution. Both methods significantly increase system overhead—for example, SeConRAG's runtime is three to four times that of the original RAG.

Another category of methods is based on the voting-consistency assumption, which posits a semantic consensus among benign documents. ReliabilityRAG \cite{shenReliabilityRAGEffectiveProvably2025} defends against attacks by constructing a contradiction graph and transforming it into a maximum independent set problem. However, as this problem is NP-hard\cite{https://doi.org/10.1002/jgt.3190170407}, processing large scale retrieval sets requires discarding a substantial number of documents to reduce computational costs, leading to significant information loss. Similarly, RobustRAG \cite{xiangCertifiablyRobustRAG2024} extracts document keywords as context, which also results in information loss. Such methods tend to fail when the density of malicious documents increases.

The third category leverages adversarial knowledge, utilizing the internal knowledge and inherent security mechanisms of LLMs. For example, AstuteRAG \cite{wangAstuteRAGOvercoming2025} dynamically integrates internal knowledge with retrieved content through heuristic selection, while InstructRAG \cite{weiInstructRAGInstructingRetrievalAugmented2024} guides the retrieval process with self-generated rationales. Both approaches depend on the intrinsic knowledge and safety mechanisms of LLMs. Other studies attempt to obtain more robust representations through metric learning \cite{xianVulnerabilityApplyingRetrievalAugmented2025}; however, the scarcity of high-quality annotated data in dynamic settings limits their practical applicability.

Overall, existing defense methods focus primarily on semantic content while overlooking the importance of ranking information. Since ranking directly quantifies document relevance and context \cite{lengBidirectionalRankingPerson2013a}, it inherently contains rich security signals. Therefore, delving into the patterns within ranking behavior is crucial for constructing a robust RAG defense framework.

\subsection{Bidirectional Search}
\label{subsec:Bidirectional Search}

Bidirectional retrieval is primarily manifested in the interactive matching between the initial input and output. In the forward interaction, the focus is on understanding the content of the input to capture its semantic proximity to the query. In the backward interaction, the output interacts inversely with the input, interpreting the context of the input based on their mutual proximity. This bidirectional concept has been widely applied across various domains. In language models, bidirectional attention captures both textual content and contextual information\cite{seoBidirectionalAttentionFlow2018}. In text retrieval, performance is enhanced by generating pseudo-documents corresponding to queries and pseudo-queries corresponding to documents\cite{wangQuery2docQueryExpansion2023}. In image retrieval, different strategies are applied to images within a query's close set (ranked by original similarity) and its far set (re-ranked using contextual similarity from k-reciprocal nearest neighbors)\cite{lengBidirectionalRankingPerson2013a,qinHelloNeighborAccurate2011a,jegouContextualDissimilarityMeasure2007,zhongRerankingPersonReidentification2017}. In recommendation systems, head ranking and tail ranking are utilized to select appropriate items for users.

Inspired by this concept, we investigates the behaviors of poisoned and benign documents under bidirectional retrieval in Section \ref{subsec:pilot-experiments}. We discovered distinct distribution patterns of poisoned and benign documents in bidirectional ranking. Based on this finding, we propose the BiRD defense method, which, for the first time, successfully applies the aforementioned concept to the RAG defense problem.
\section{Motivation}

This section aims to provide a clear understanding of the behavioral differences between poisoned and benign documents in the bidirectional ranking. We first present the pilot experiments in Section~\ref{subsec:pilot-experiments} that reveal these differences, followed by a reason analysis in Section~\ref{subsec:reason analysis} that explains the underlying mechanisms.
\subsection{Pilot Experiments}
\label{subsec:pilot-experiments}
To gain an in-depth understanding of the behavioral differences between poisoned and benign documents in the retrieval system, we conducted statistical analyses of bidirectional ranking patterns using the HotpotQA, MS-MARCO, NQ, and PROCON datasets, along with the Contriever, DPR, and ANCE retrievers, under PoisonedRAG\cite{zouPoisonedRAGKnowledgeCorruption2024} and Topic-FlipRAG\cite{gongTopicFlipRAGTopicOrientatedAdversarial2025} attack. Given the massive scale of the full HotpotQA, MS-MARCO, and NQ datasets (each containing millions of documents), we first constructed a subset. For each target query, we selected the 20 most relevant benign documents from the corpus and additionally poisoned 5 documents to form the subset. We then performed forward retrieval on the three retrievers to obtain the top 20 most relevant documents. The frequency of poisoned documents appearing at each rank position constitutes the first column of the heatmap. For each document retrieved in the forward ranking, we conducted a backward retrieval to obtain its backward ranking. We then calculated the frequency of poisoned documents at each rank position in these backward rankings. Each backward ranking corresponds to one row in the right side of heatmap, with the k-th row representing the backward ranking result of the document that ranked k-th in the forward ranking.

Figure \ref{fig:combined_results} shows the distribution of poisoned documents in the bidirectional ranking under PoisonedRAG attack on the HotpotQA dataset with different retrievers. Figure \ref{fig:combined_results} shows the distribution of poisoned documents in the bidirectional ranking under Topic-FlipRAG attack on the PROCON datasets. More details can be found in the Appendix \ref{app:preliminary_experiments}.

Through systematic experimental analysis as shown in Figures \ref{fig:combined_results}, we have discovered and revealed for the first time systematic differences in backward ranking behavior between poisoned and benign documents:
\begin{figure}[!t]
\centering
\includegraphics[width=\columnwidth]{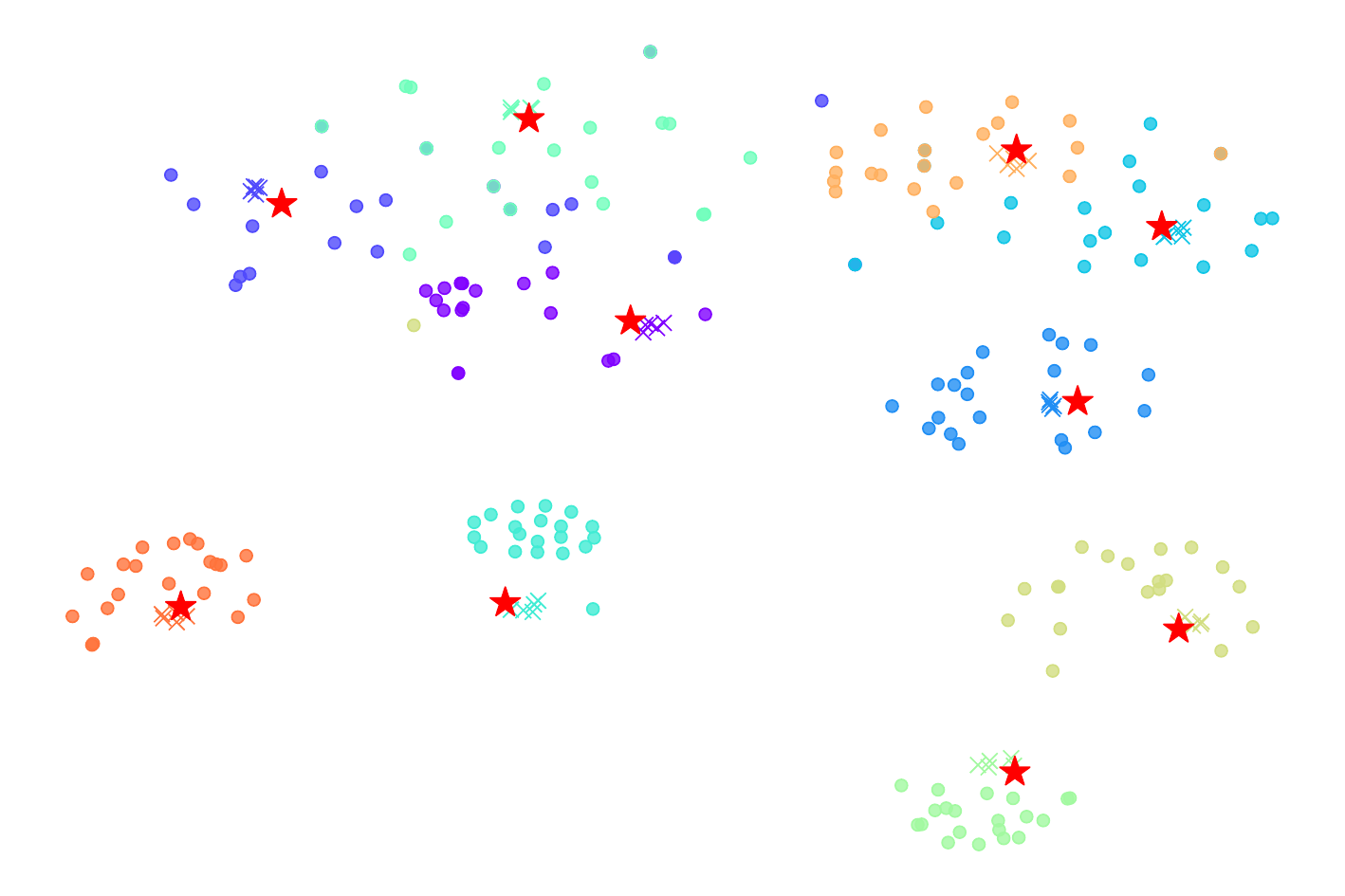}
\caption{t-SNE visualization of textual embeddings for benign documents, poisoned documents, and queries generated by Contriever on the subset of HotpotQA.}
\label{fig:tsne_visualization}
\end{figure}
\begin{figure*}[ht]
 \centering 
 \includegraphics[width=\textwidth]{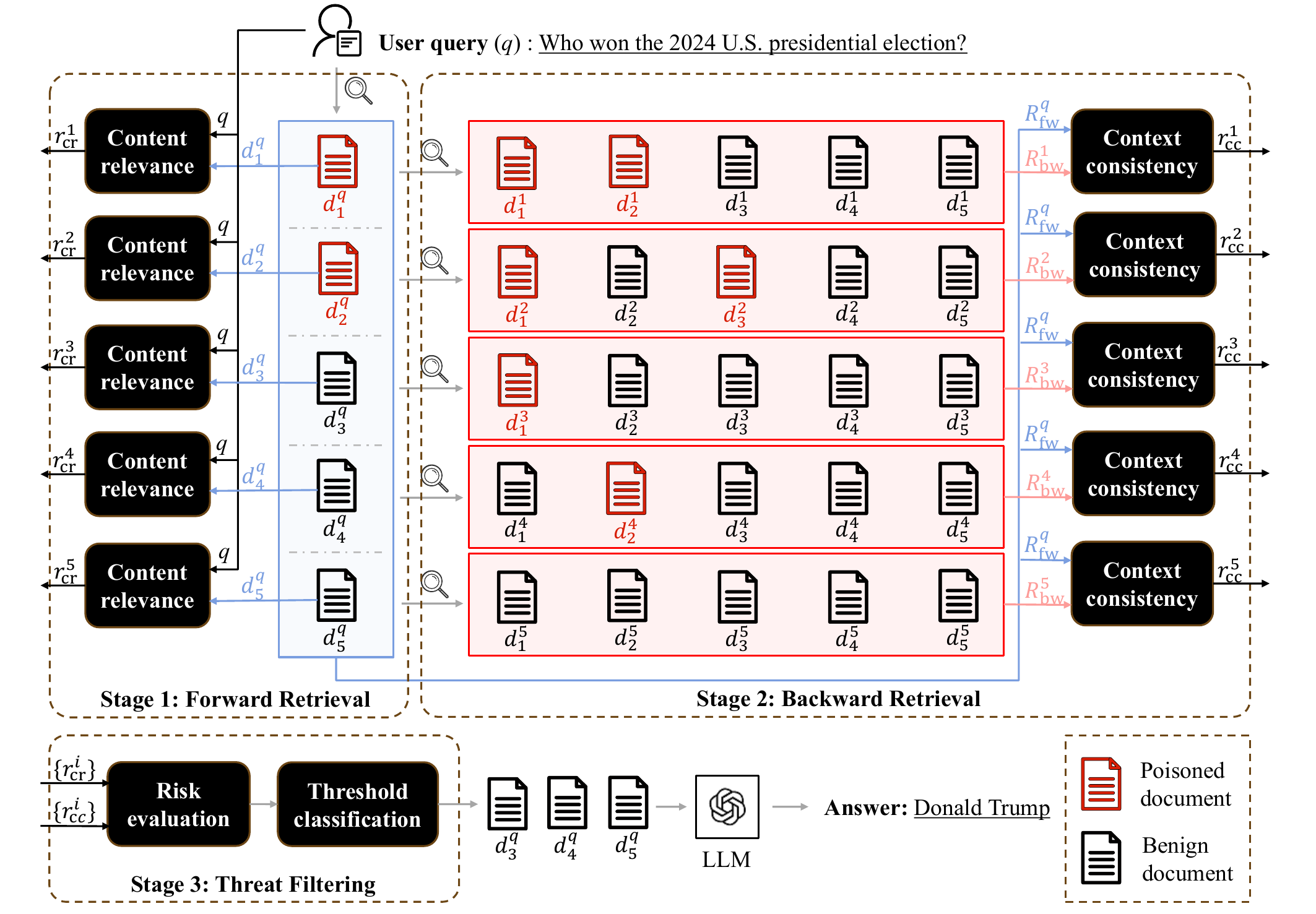}
\caption{Overview of BiRD method. The process is mainly divided into three stages: forward retrieval, backward retrieval, and threat filtering. It integrates the query-document content relevance from the forward ranking and context consistency from the bidirectional ranking to detect and filter poisoned documents.}
 \label{fig:pipeline}
 \end{figure*}
\begin{itemize} \item \textbf{Forward Ranking Distribution Asymmetry:} In the forward ranking process, poisoned documents exhibit a "top-heavy" distribution, predominantly concentrating at the peak of the retrieval list to maximize their influence. In contrast, benign documents are largely dispersed toward the bottom, failing to achieve a comparable level of retrieval priority. \item \textbf{Backward Ranking Stability:} When generating a backward ranking using a poisoned document as the pivot, other poisoned documents remain highly concentrated at the top positions, reflecting the strong contextual correlation among them. Conversely, the backward ranking of a benign document shows a significantly more uniform and consistent distribution across the corpus, reflecting the inherent diversity of benign content. \item \textbf{Forward-Backward Consistency:} Poisoned documents exhibit a high degree of structural similarity between the forward ranking of queries and their own backward ranking. In contrast, benign documents manifest a significant divergence between the two, suggesting a more complex and diverse set of contextual relationships.\end{itemize}

\subsection{Reason Analysis}
\label{subsec:reason analysis}

To further elucidate the underlying mechanisms governing our statistical observations, we visualize, using t-SNE, the textual embeddings generated by Contriever for the subset of the HotpotQA dataset that we have constructed. As illustrated in Figure \ref{fig:tsne_visualization}, the visualization demonstrates a stark dichotomy in the distribution of the two document types. The poisoned documents (red crosses) tend to form highly dense clusters. In stark contrast, the embeddings of benign documents (green circles) are widely and uniformly dispersed, reflecting the inherent semantic diversity of a genuine knowledge base.

While prior research has observed the tendency of poisoned documents to cluster \cite{zhouTrustRAGEnhancingRobustness2025}, a reasonable explanation has been lacking. As shown in the projection, to maximize the attack success rate and ensure that the LLM generates the target adversarial output with high confidence, attackers must elevate the ranking position (determined by semantic similarity) of otherwise irrelevant poisoned documents. To achieve this, attackers employ various techniques to optimize the embeddings of poisoned documents, bringing them closer to the target query. Simultaneously, they must constrain the magnitude of these modifications to avoid easy detection. Consequently, these documents approach the query embedding along similar trajectories, causing the generated malicious documents to become highly consistent in the embedding space and converge into tight clusters. In contrast, benign documents, due to their diverse origins and heterogeneous subject matter, remain naturally scattered in a diffuse distribution.

Based on this observation, our core argument is as follows: poisoned documents can be effectively distinguished and filtered by analyzing content relevance and context consistency to the query. content relevance, determined by semantic similarity, is informed by the finding of forward ranking distribution asymmetry. context consistency, reflecting a document's context consistency to the query through ranking similarity between forward and backward rankings, is informed by the findings of backward ranking stability and forward-backward consistency. The primary advantage of BiRD is that it does not rely on LLMs to extract deep semantic information; instead, it fully leverages the ranking signals inherent in the retrieval process and achieves lower computation cost, distilling robust security features from the system's fundamental ranking behavior.
\section{Methodology}
\label{sec:methodology}

This section details the BiRD method as illustrated in Figure \ref{fig:pipeline}. First, Section~\ref{subsec:problem-definition} formally defines the problem to be addressed in RAG security. Subsequently, we introduces BiRD method, including forward retrieval (Section~\ref{subsec:forward retrieval}), backward retrieval (Section~\ref{subsec:backward retrieval}), and threat filtering (Section~\ref{subsec:threat filtering}). Finally, we analyze the time and space complexity of BiRD in Section~\ref{subsec:time-and-space-complexity}.
\subsection{Threat Model}
\label{subsec:problem-definition}

We consider a poisoning attack scenario where an adversary aims to mislead a RAG system by injecting carefully crafted malicious documents into the retrieval corpus. The attacker selects a set of target queries $\mathcal{Q}_{\text{target}}$ and specifies an intended false answer $r(q)$ for each $q \in \mathcal{Q}_{\text{target}}$. For instance, for a query such as ``Who won the 2024 U.S. presidential election?'', the attacker may intend the RAG system to output the wrong answer $r(q)=\text{``Kamala Harris''}$.

To achieve this goal, the attacker injects a set of poisoned documents $\mathcal{D}_{\text{poison}}$ into the corpus. We denote the original corpus as $\mathcal{D}_{\text{benign}}$, and the poisoned corpus as
\begin{equation}
\label{eq:poisoned-corpus}
\mathcal{D}' = \mathcal{D}_{\text{benign}} \cup \mathcal{D}_{\text{poison}}.
\end{equation}

The objective of the corpus poisoning attack is that for a target query $q \in \mathcal{Q}_{\text{target}}$, at least one poisoned document is retrieved in the top-$k$ results from $\mathcal{D}'$. Formally, let
\begin{equation}
\label{eq:topk-threat}
\text{top-}k\bigl(R(q, \mathcal{D}')\bigr) \leftarrow d_1 < d_2 < \cdots < d_k,
\end{equation}
where $R(\cdot,\cdot)$ is the retriever and $d_1 < \cdots < d_k$ denotes the ranking order. 

Subsequently, these $k$ documents serve as context, along with the query, as input to a LLM. The attack is considered successful if the LLM produce the target false answer:
\begin{equation}
\label{eq:attack-goal}
\text{LLM}\bigl(q, \text{top-}k(R(q, \mathcal{D}'))\bigr) = r(q), \quad \forall q \in \mathcal{Q}_{\text{target}}.
\end{equation}

From the defender's perspective, we operates without any assumptions regarding the attack strategy, making it applicable to both white-box and black-box scenarios. The defender only has access to the initial ranking list returned by the retriever ${R}$ for a given user query $q$, which may contains a mixture of benign and poisoned documents. The defender's objective is to filter out the poisoned documents from this initial ranking without knowing which documents are malicious, thereby obtaining a clean subset of documents $\mathcal{D}_\text{clean}$, where $|\mathcal{D}_\text{clean}|=N $ . Ideally, the filtered context provided to the LLM should contain no poisoned documents.

\subsection{Forward Retrieval}
\label{subsec:forward retrieval}
\textbf{Content relevance:} To quantitatively measure the semantic content relevance, we utilize the semantic similarity function provided by the retriever. Given a user query $q$ and a corpus $\mathcal{D}'$, the retriever ${R}$ returns the top-$k$ documents based on their relevance to the query, forming the forward ranking (an ordered list):
\begin{equation}
\label{eq:forward-ranking}
R_{\text{fw}}^q \leftarrow \text{top-}k\bigl(R(q, \mathcal{D}')\bigr) \leftarrow d_1^q < d_2^q < \cdots < d_k^q.
\end{equation}
Here, $d_i^q$ denotes the $i$-th document in the forward ranking of query $q$. For each $d_i^q$ in this ranking, its content relevance is computed as:
\begin{equation}
\label{eq:rcr}
r_{cr}^i = \text{sim}(q, d_i^q),
\end{equation}
where $\text{sim}(\cdot, \cdot)$ denotes the similarity function used by the retriever, e.g., cosine similarity.

\subsection{Backward Retrieval}
\label{subsec:backward retrieval} 
For each document $d_i^q$ in the forward ranking $R_{\text{fw}}^q$, we treat $d_i^q$ itself as a new query. We then perform retrieval on the corpus to obtain its backward ranking (also an ordered list):
\begin{equation}
\label{eq:backward-ranking}
R_{\text{bw}}^i \leftarrow \text{top-}k\bigl(R(d_i^q, \mathcal{D}')\bigr) \leftarrow d_1^i < d_2^i < \cdots < d_k^i.
\end{equation}
Here, $d_j^i$ denotes the $j$-th document in the backward ranking of query $d_i^q$.
\textbf{Context consistency:} To measure the ranking context consistency, we compute the ranking similarity between the backward ranking and the forward ranking:
\begin{equation}
\label{eq:sbw}
\begin{aligned}
r_{cc}^i &= \varrho\left( R_{\text{fw}}^q, R_{\text{bw}}^i \right)\\
&=
\begin{cases}
1 - \dfrac{6 \sum\limits_{d \in \mathcal{C}} \left(r_f(d) - r_b(d)\right)^2}{\begin{array}{l}
|\mathcal{C}|\bigl(|\mathcal{C}|^2 - 1\bigr)\\
\end{array}}, & \text{if } |\mathcal{C}| \geq 2, \\
0, & \text{if } |\mathcal{C}| < 2.
\end{cases}
\end{aligned}
\end{equation}
where
\begin{equation}
\label{eq:common-set}
\mathcal{C} = \{ d \mid d \in R_{\text{fw}}^q \ \wedge\  d \in R_{\text{bw}}^i \}
\end{equation}
represents the set of common documents in both rankings, $|\mathcal{C}|$ denotes the number of common documents, and $r_f(d)$ and $r_b(d)$ denote the rank of document $d$ in $R_{\text{fw}}^q$ and $R_{\text{bw}}^i$, respectively. $r_{cc}^i \in [-1,1]$. According to the ranking inequality, when the forward ranking $R_{\text{fw}}^q$ and the backward ranking $R_{\text{bw}}^i$ completely coincide, $r_{cc}^i = 1$; when they are completely opposite, $r_{cc}^i = -1$. Here, $\varrho$ represents the Spearman's rank correlation coefficient. We will discuss the impact of ranking similarity method in detail in Section~\ref{subsec:dynamic-performance-experiments}.

\subsection{Threat Filtering}
\label{subsec:threat filtering}
\textbf{Risk evaluation:}Based on the key observation that poisoned documents tend to exhibit high backward/forward ranking similarity (i.e., $r_{cc}^i$ is close to 1), whereas the backward ranking of a benign document significantly diverges from the forward ranking (i.e., for each $d \in \mathcal{D}_{\text{benign}}$, $r_{cc}$ is close to 0), we design the following filtering criterion.

For each document $d_i^q \in R_{\text{fw}}^q$, to amplify the distinction between poisoned and benign documents, we compute a final composite score using division:
\begin{equation}
\label{eq:composite-score}
S(d_i^q) = \frac{r_{cr}^i}{1 - r_{cc}^i}.
\end{equation}
\textbf{Threshold classification:} Given a predefined threshold $\epsilon$, the final set of benign documents $\mathcal{D}_{\text{clean}}$ is then:
\begin{equation}
\label{eq:clean-set}
\mathcal{D}_{\text{clean}} = \{ d_i^q \mid d_i^q \in R_{\text{fw}}^q \text{ and } S(d_i^q) \leq \epsilon \}.
\end{equation}

In our method, the threshold $\epsilon$ is currently selected empirically. In practical deployment, $\epsilon$ can be determined by sampling poisoned documents generated by existing attack methods to form a validation set (see details in Section \ref{subsec:validation-threshold-selection}). The impact of the threshold $\epsilon$ will be discussed in detail in Section~\ref{subsec:dynamic-performance-experiments}.
The complete algorithm is detaild in Algorithm~\ref{alg:algorithm1}.

\begin{algorithm}[!t]
\caption{Bidirectional Ranking Defense (BiRD)}
\label{alg:algorithm1}
\begin{algorithmic}[1]
\Require User query $q$, document corpus ${D}$, retriever ${R}$, parameters $k$, $\epsilon$
\Ensure Benign document set $\mathcal{D}_{\text{clean}}$
\State $R_{\text{fw}}^q \gets {R}(q, \mathcal{D}')$ \Comment{Forward ranking}
\For{each document $d_i^q \in R_{\text{fw}}^q$}
    \State $R_{\text{bw}}^i \gets {R}(d_i^q, \mathcal{D}' \setminus \{d_i^q\})$ \Comment{Backward ranking}
    \State $r_{cr}^i \gets \text{sim}(q, d_i^q)$ 
    \State $r_{cc}^i \gets \varrho(R_{\text{fw}}^q, R_{\text{bw}}^i)$ 
    \State $S(d_i^q) \gets r_{cr}^i / (1 - r_{cc}^i)$
\EndFor
\State $\mathcal{D}_{\text{clean}} \gets \{ d_i^q \mid d_i^q \in R_{\text{fw}}^q \text{ and } S(d_i^q) \leq \epsilon \}$ \Comment{Filter to obtain clean set}
\State \Return $\mathcal{D}_{\text{clean}}$
\end{algorithmic}
\end{algorithm}
\subsection{Complexity Analysis}
\label{subsec:time-and-space-complexity}
\textbf{Time complexity}: Assume that the time complexity of a single retrieval operation by the retriever ${R}$ is $O(N)$, where $N = |\mathcal\mathcal{D}'|$ is the total number of documents in the corpus. The overall time complexity of the BiRD algorithm consists of the following components: one forward retrieval $O(N)$, $k$ backward retrievals $O(kN)$, and the computation of ranking correlations $O(k^2)$ (e.g., using the Spearman's rank correlation coefficient). Therefore, the total time complexity is $O((k+1)N + k^2)$. Since typically $k \ll N$ (e.g., $k=20$, while $N$ can be on the order of millions), and if matrix computation is employed, the $k$ backward retrievals can be performed simultaneously, then the dominant overhead stems from approximately 2 times the retrieval operation time (i.e., one forward and one effectively parallelized backward retrieval). In practice, the retrieval time in RAG often constitutes only a small fraction of the overall RAG pipeline. Compared to other methods that require invoking external LLMs for generation, our approach essentially does not affect the real-world user experience of RAG.

\textbf{Space complexity:} The space complexity of BiRD is O($N$), consistent with the original retriever. If one opts to precompute and store the
dense pairwise similarity matrix, the space overhead increases to O($N^2$).

Furthermore, in practical applications, the document corpus is usually static and known in advance. Thus, pairwise similarities between all documents can be precomputed and cached. This allows the reuse of existing similarity results during the backward retrieval phase, avoiding redundant calculations. This precomputation strategy significantly reduces the runtime overhead of the algorithm. 
\FloatBarrier
\begin{table*}[!t]
\centering
\caption{Experimental Results on NQ. Best results ($\downarrow$ for ASR, $\uparrow$ for ACC) are in bold; second-best results are underlined.}
\label{tab:nq}
\fontsize{8}{10}\selectfont
\setlength{\tabcolsep}{1.2pt}
\begin{tabular*}{\textwidth}{@{\extracolsep{\fill}} ccc *{9}{c} @{}}
\toprule
\multirow{2.5}{*}{\textbf{Model}} & \multirow{2.5}{*}{\textbf{Baseline}} & \multirow{2.5}{*}{\textbf{Venue}} & \multicolumn{3}{c}{\textbf{ANCE}} & \multicolumn{3}{c}{\textbf{Contriever}} & \multicolumn{3}{c}{\textbf{DPR}} \\
\cmidrule{4-6} \cmidrule{7-9} \cmidrule{10-12}
& & & {Clean} & {PIA} & {PoisonedRAG} & {Clean} & {PIA} & {PoisonedRAG} & {Clean} & {PIA} & {PoisonedRAG} \\
\midrule
\multirow{5}{*}{\textbf{Qwen2.5-7B}} & VanillaRAG & {-} & \textbf{0.76} & \textbf{0.20} / \underline{0.59} & 0.87 / 0.37 & \textbf{0.77} & \underline{0.26} / \underline{0.69} & 0.89 / 0.34 & \textbf{0.73} & \underline{0.29} / \underline{0.58} & 0.87 / \underline{0.44} \\
& RobustRAG & arXiv '24 & 0.60 & 0.74 / 0.48 & \underline{0.61} / 0.40 & 0.62 & 0.79 / 0.49 & \underline{0.62} / 0.42 & 0.60 & 0.70 / 0.50 & \underline{0.62} / 0.43 \\
& InstructRAG & ICLR '25 & \underline{0.68} & 0.55 / 0.49 & 0.77 / 0.29 & \underline{0.63} & 0.51 / 0.57 & 0.79 / 0.24 & 0.63 & 0.54 / 0.48 & 0.79 / 0.32 \\
& ReliabilityRAG & NeurIPS '25 & 0.48 & 0.41 / 0.44 & \underline{0.61} / \underline{0.43} & 0.54 & 0.40 / 0.44 & 0.70 / \underline{0.53} & 0.55 & 0.32 / 0.50 & 0.70 / 0.41 \\
& BiRD & Ours & 0.61 & \underline{0.22} / \textbf{0.64} & \textbf{0.08} / \textbf{0.75} & 0.62 & \textbf{0.25} / \textbf{0.71} & \textbf{0.16} / \textbf{0.62} & \underline{0.64} & \textbf{0.25} / \textbf{0.60} & \textbf{0.30} / \textbf{0.63} \\
\midrule
\multirow{5}{*}{\textbf{Mistral-7B}} & VanillaRAG & {-} & \textbf{0.77} & \textbf{0.16} / \underline{0.60} & 0.93 / 0.16 & \textbf{0.76} & \underline{0.20} / \underline{0.59} & 0.91 / 0.20 & \textbf{0.80} & \underline{0.29} / \textbf{0.56} & 0.91 / 0.18 \\
& RobustRAG & arXiv '24 & 0.21 & 0.33 / 0.15 & \underline{0.40} / 0.17 & 0.17 & 0.33 / 0.14 & 0.49 / 0.18 & 0.16 & 0.34 / 0.18 & 0.45 / 0.14 \\
& InstructRAG & ICLR '25 & 0.38 & 0.48 / 0.39 & 0.63 / \underline{0.27} & 0.25 & 0.39 / 0.51 & 0.54 / \underline{0.33} & 0.30 & 0.34 / 0.48 & 0.55 / \underline{0.31} \\
& ReliabilityRAG & NeurIPS '25 & 0.12 & 0.29 / 0.17 & \underline{0.40} / 0.18 & 0.09 & \textbf{0.18} / 0.15 & \underline{0.35} / 0.14 & 0.12 & \textbf{0.28} / 0.11 & \underline{0.41} / 0.15 \\
& BiRD & Ours & \underline{0.62} & \underline{0.19} / \textbf{0.61} & \textbf{0.06} / \textbf{0.71} & \underline{0.65} & \underline{0.20} / \textbf{0.64} & \textbf{0.10} / \textbf{0.72} & \underline{0.66} & \underline{0.29} / \underline{0.55} & \textbf{0.24} / \textbf{0.65} \\
\midrule
\multirow{5}{*}{\textbf{Llama-3.1-8B}} & VanillaRAG & {-} & \textbf{0.88} & \underline{0.21} / \underline{0.72} & 0.83 / \underline{0.78} & \textbf{0.87} & \underline{0.30} / 0.77 & 0.85 / \underline{0.76} & \textbf{0.89} & \textbf{0.31} / 0.65 & 0.84 / \underline{0.68} \\
& RobustRAG & arXiv '24 & 0.36 & 0.75 / 0.52 & \underline{0.50} / 0.45 & 0.30 & 0.79 / 0.57 & \underline{0.45} / 0.33 & 0.25 & 0.72 / 0.59 & \underline{0.51} / 0.44 \\
& InstructRAG & ICLR '25 & \underline{0.83} & 0.34 / \textbf{0.81} & 0.59 / 0.67 & \underline{0.83} & 0.31 / \textbf{0.81} & 0.57 / 0.68 & \underline{0.78} & \underline{0.34} / \textbf{0.75} & 0.69 / 0.62 \\
& ReliabilityRAG & NeurIPS '25 & 0.22 & 0.68 / 0.44 & \underline{0.50} / 0.39 & 0.31 & 0.64 / 0.44 & 0.46 / 0.33 & 0.24 & 0.64 / 0.39 & 0.52 / 0.42 \\
& BiRD & Ours & 0.72 & \textbf{0.18} / \underline{0.72} & \textbf{0.15} / \textbf{0.79} & 0.79 & \textbf{0.29} / \underline{0.80} & \textbf{0.19} / \textbf{0.82} & 0.77 & 0.38 / \underline{0.66} & \textbf{0.28} / \textbf{0.75} \\
\bottomrule
\end{tabular*}
\end{table*}

\begin{table*}[!t]
\centering
\caption{Experimental Results on Msmarco. Best results ($\downarrow$ for ASR, $\uparrow$ for ACC) are in bold; second-best results are underlined.}
\label{tab:msmarco}
\fontsize{8}{10}\selectfont
\setlength{\tabcolsep}{1.2pt}
\begin{tabular*}{\textwidth}{@{\extracolsep{\fill}} ccc *{9}{c} @{}}
\toprule
\multirow{2.5}{*}{\textbf{Model}} & \multirow{2.5}{*}{\textbf{Baseline}} & \multirow{2.5}{*}{\textbf{Venue}} & \multicolumn{3}{c}{\textbf{ANCE}} & \multicolumn{3}{c}{\textbf{Contriever}} & \multicolumn{3}{c}{\textbf{DPR}} \\
\cmidrule{4-6} \cmidrule{7-9} \cmidrule{10-12}
& & & {Clean} & {PIA} & {PoisonedRAG} & {Clean} & {PIA} & {PoisonedRAG} & {Clean} & {PIA} & {PoisonedRAG} \\
\midrule
\multirow{5}{*}{\textbf{Qwen2.5-7B}} & VanillaRAG & {-} & \textbf{0.81} & \underline{0.26} / \textbf{0.69} & 0.86 / 0.47 & \textbf{0.81} & \textbf{0.26} / \textbf{0.63} & 0.85 / 0.49 & \textbf{0.78} & \underline{0.26} / \textbf{0.65} & 0.88 / 0.47 \\
& RobustRAG & arXiv '24 & 0.71 & 0.69 / 0.52 & 0.69 / 0.56 & 0.66 & 0.71 / 0.61 & 0.69 / 0.48 & 0.69 & 0.65 / 0.58 & 0.67 / 0.51 \\
& InstructRAG & ICLR '25 & \underline{0.79} & 0.63 / 0.56 & 0.80 / 0.35 & \underline{0.76} & 0.53 / 0.55 & 0.84 / 0.35 & \underline{0.75} & 0.50 / 0.55 & 0.79 / 0.32 \\
& ReliabilityRAG & NeurIPS '25 & 0.64 & 0.27 / \underline{0.59} & \underline{0.58} / \underline{0.60} & 0.64 & 0.30 / \underline{0.62} & \underline{0.57} / \underline{0.55} & 0.70 & 0.31 / 0.59 & \underline{0.58} / \underline{0.57} \\
& BiRD & Ours & 0.59 & \textbf{0.22} / \textbf{0.69} & \textbf{0.17} / \textbf{0.76} & 0.60 & \underline{0.27} / \textbf{0.63} & \textbf{0.24} / \textbf{0.71} & 0.64 & \textbf{0.25} / \underline{0.63} & \textbf{0.27} / \textbf{0.68} \\
\midrule
\multirow{5}{*}{\textbf{Mistral-7B}} & VanillaRAG & {-} & \textbf{0.81} & 0.13 / \textbf{0.72} & 0.88 / \underline{0.24} & \textbf{0.86} & 0.20 / \underline{0.59} & 0.75 / \underline{0.32} & \textbf{0.77} & 0.23 / \underline{0.55} & 0.76 / \underline{0.32} \\
& RobustRAG & arXiv '24 & 0.20 & \underline{0.04} / 0.04 & 0.22 / 0.08 & 0.24 & \underline{0.04} / 0.02 & 0.28 / 0.11 & 0.23 & \textbf{0.03} / 0.08 & 0.18 / 0.13 \\
& InstructRAG & ICLR '25 & 0.24 & 0.14 / 0.57 & 0.30 / 0.21 & 0.25 & 0.15 / 0.51 & 0.28 / \underline{0.32} & 0.25 & \underline{0.15} / 0.49 & 0.37 / 0.28 \\
& ReliabilityRAG & NeurIPS '25 & 0.18 & \textbf{0.03} / 0.06 & \underline{0.20} / 0.05 & 0.18 & \textbf{0.03} / 0.06 & \underline{0.19} / 0.16 & 0.12 & \textbf{0.03} / 0.11 & \underline{0.17} / 0.12 \\
& BiRD & Ours & \underline{0.69} & 0.12 / \underline{0.69} & \textbf{0.12} / \textbf{0.75} & \underline{0.71} & 0.23 / \textbf{0.62} & \textbf{0.14} / \textbf{0.76} & \underline{0.67} & 0.20 / \textbf{0.61} & \textbf{0.14} / \textbf{0.71} \\
\midrule
\multirow{5}{*}{\textbf{Llama-3.1-8B}} & VanillaRAG & {-} & \textbf{0.90} & \underline{0.29} / \textbf{0.83} & 0.82 / \underline{0.77} & \textbf{0.90} & \textbf{0.35} / \underline{0.82} & 0.89 / \underline{0.74} & \textbf{0.89} & \textbf{0.31} / 0.73 & 0.81 / \underline{0.79} \\
& RobustRAG & arXiv '24 & 0.34 & 0.81 / 0.61 & \underline{0.39} / 0.50 & 0.40 & 0.72 / 0.54 & 0.39 / 0.49 & 0.35 & 0.59 / 0.59 & 0.43 / 0.51 \\
& InstructRAG & ICLR '25 & 0.76 & 0.39 / 0.72 & 0.60 / 0.70 & 0.74 & 0.37 / 0.74 & 0.58 / \underline{0.74} & 0.74 & 0.49 / \underline{0.75} & 0.61 / 0.69 \\
& ReliabilityRAG & NeurIPS '25 & 0.40 & 0.67 / 0.48 & 0.46 / 0.44 & 0.39 & 0.65 / 0.54 & \textbf{0.33} / 0.34 & 0.30 & 0.56 / 0.53 & \underline{0.37} / 0.39 \\
& BiRD & Ours & \underline{0.82} & \textbf{0.28} / \underline{0.82} & \textbf{0.30} / \textbf{0.87} & \underline{0.81} & \underline{0.36} / \textbf{0.83} & \underline{0.35} / \textbf{0.87} & \underline{0.83} & \underline{0.35} / \textbf{0.76} & \textbf{0.36} / \textbf{0.83} \\
\bottomrule
\end{tabular*}
\end{table*}

\begin{table*}[!t]
\centering
\caption{Experimental Results on HotpotQA. Best results ($\downarrow$ for ASR, $\uparrow$ for ACC) are in bold; second-best results are underlined.}
\label{tab:hotpotqa}
\fontsize{8}{10}\selectfont
\setlength{\tabcolsep}{1.2pt}
\begin{tabular*}{\textwidth}{@{\extracolsep{\fill}} ccc *{9}{c} @{}}
\toprule
\multirow{2.5}{*}{\textbf{Model}} & \multirow{2.5}{*}{\textbf{Baseline}} & \multirow{2.5}{*}{\textbf{Venue}} & \multicolumn{3}{c}{\textbf{ANCE}} & \multicolumn{3}{c}{\textbf{Contriever}} & \multicolumn{3}{c}{\textbf{DPR}} \\
\cmidrule{4-6} \cmidrule{7-9} \cmidrule{10-12}
& & & {Clean} & {PIA} & {PoisonedRAG} & {Clean} & {PIA} & {PoisonedRAG} & {Clean} & {PIA} & {PoisonedRAG} \\
\midrule
\multirow{5}{*}{\textbf{Qwen2.5-7B}} & VanillaRAG & {-} & \textbf{0.61} & \textbf{0.34} / \underline{0.49} & 0.93 / 0.24 & \underline{0.66} & \textbf{0.32} / 0.54 & 0.93 / 0.29 & \textbf{0.66} & \textbf{0.42} / \textbf{0.55} & 0.95 / 0.24 \\
& RobustRAG & arXiv '24 & \underline{0.54} & 0.74 / \textbf{0.50} & 0.88 / 0.33 & 0.58 & 0.73 / 0.52 & 0.79 / \underline{0.41} & 0.59 & 0.65 / 0.41 & 0.78 / 0.30 \\
& InstructRAG & ICLR '25 & \underline{0.54} & 0.60 / \underline{0.49} & 0.81 / 0.20 & \textbf{0.67} & 0.57 / \underline{0.59} & 0.82 / 0.16 & 0.59 & 0.63 / 0.53 & 0.84 / 0.19 \\
& ReliabilityRAG & NeurIPS '25 & 0.48 & \underline{0.42} / 0.33 & \underline{0.65} / \underline{0.36} & 0.50 & \underline{0.36} / 0.39 & \underline{0.70} / 0.39 & 0.42 & \textbf{0.42} / 0.39 & \underline{0.60} / \underline{0.31} \\
& BiRD & Ours & 0.49 & \textbf{0.34} / \underline{0.49} & \textbf{0.22} / \textbf{0.59} & 0.57 & \underline{0.36} / \textbf{0.60} & \textbf{0.24} / \textbf{0.69} & \underline{0.61} & \underline{0.48} / \underline{0.54} & \textbf{0.41} / \textbf{0.62} \\
\midrule
\multirow{5}{*}{\textbf{Mistral-7B}} & VanillaRAG & {-} & \textbf{0.74} & \underline{0.27} / \textbf{0.49} & 0.93 / 0.15 & \textbf{0.79} & 0.27 / \textbf{0.67} & 0.93 / 0.16 & \textbf{0.72} & 0.39 / \underline{0.46} & 0.95 / 0.13 \\
& RobustRAG & arXiv '24 & 0.08 & \textbf{0.19} / 0.04 & 0.41 / 0.11 & 0.02 & \underline{0.22} / 0.06 & \underline{0.29} / 0.10 & 0.04 & \underline{0.17} / 0.05 & \underline{0.35} / 0.12 \\
& InstructRAG & ICLR '25 & \underline{0.65} & 0.67 / 0.29 & 0.78 / \underline{0.24} & \underline{0.74} & 0.57 / 0.40 & 0.80 / \underline{0.28} & 0.60 & 0.65 / 0.29 & 0.89 / \underline{0.19} \\
& ReliabilityRAG & NeurIPS '25 & 0.06 & \textbf{0.19} / 0.04 & \underline{0.33} / 0.12 & 0.04 & \textbf{0.18} / 0.04 & 0.36 / 0.13 & 0.02 & \textbf{0.13} / 0.04 & \underline{0.35} / 0.11 \\
& BiRD & Ours & 0.61 & \underline{0.27} / \underline{0.46} & \textbf{0.18} / \textbf{0.67} & 0.69 & 0.27 / \underline{0.59} & \textbf{0.17} / \textbf{0.68} & \underline{0.69} & 0.38 / \textbf{0.48} & \textbf{0.28} / \textbf{0.59} \\
\midrule
\multirow{5}{*}{\textbf{Llama-3.1-8B}} & VanillaRAG & {-} & \textbf{0.79} & \textbf{0.35} / \underline{0.60} & 0.88 / \underline{0.65} & \textbf{0.87} & \textbf{0.33} / \underline{0.69} & 0.79 / \underline{0.71} & \textbf{0.75} & \underline{0.43} / \underline{0.56} & 0.90 / \underline{0.61} \\
& RobustRAG & arXiv '24 & 0.21 & 0.73 / 0.42 & 0.44 / 0.37 & 0.22 & 0.78 / 0.52 & 0.45 / 0.37 & 0.26 & 0.70 / 0.42 & \underline{0.51} / 0.38 \\
& InstructRAG & ICLR '25 & \underline{0.78} & 0.53 / 0.50 & 0.79 / 0.42 & \underline{0.82} & 0.44 / 0.42 & 0.61 / 0.51 & \underline{0.69} & \textbf{0.38} / 0.35 & 0.68 / 0.41 \\
& ReliabilityRAG & NeurIPS '25 & 0.21 & 0.49 / 0.31 & \underline{0.36} / 0.30 & 0.26 & 0.51 / 0.40 & \underline{0.38} / 0.29 & 0.22 & 0.60 / 0.33 & \textbf{0.44} / 0.26 \\
& BiRD & Ours & 0.70 & \underline{0.39} / \textbf{0.64} & \textbf{0.33} / \textbf{0.82} & 0.78 & \underline{0.37} / \textbf{0.72} & \textbf{0.33} / \textbf{0.85} & \textbf{0.75} & 0.50 / \textbf{0.64} & \underline{0.51} / \textbf{0.78} \\
\bottomrule
\end{tabular*}
\end{table*}

\section{Experiments}
This chapter provides a detailed description of the extensive and systematic experiments conducted to evaluate our defense method. We first introduce the fundamental experimental setup (Section \ref{subsec:experimental-setup}), followed by a comprehensive discussion of the baseline experiments (Section \ref{subsec:baseline-experiments}), dynamic performance experiments (Section \ref{subsec:dynamic-performance-experiments}) , and runtime experiments (Section \ref{subsec:runtime-experiments}).
\subsection{Experimental Setup} 
\label{subsec:experimental-setup}
\textbf{Datasets:} Our evaluation encompasses three prominent and widely-adopted open-domain question answering datasets, chosen for their diversity in complexity and task nature: Natural Questions (NQ)\cite{kwiatkowskiNaturalQuestionsBenchmark2019}, which consists of real-world questions derived from Google search queries; MSMARCO\cite{bajajMSMARCOHuman2018}, a large-scale machine reading comprehension dataset from Microsoft; and HotpotQA\cite{yangHotpotQADatasetDiverse2018b}, a challenging dataset designed for multi-hop reasoning. These datasets collectively cover a spectrum from straightforward factoid retrieval to complex, interconnected reasoning scenarios, thereby enabling a robust and holistic assessment of our defense method's performance and resilience across varied contexts.

\textbf{Retrievers:} To ensure the general applicability of our approach across different retrieval paradigms, we tested our defense method using three prevalent and widely-used dense retrieval models: Contriever \cite{izacardUnsupervisedDenseInformation2022}, a general-purpose retrieval model trained with contrastive learning; ANCE\cite{xiongApproximateNearestNeighbor2020a} (Approximate Nearest Neighbor Negative Contrastive Estimation), which employs hard negative sampling via an approximate nearest neighbor search; and DPR\cite{karpukhinDensePassageRetrieval2020a} (Dense Passage Retriever), the seminal and commonly adopted model for dense passage retrieval.

\textbf{Large Language Models:} To validate the generalization capability of our defense method and ensure its robustness is not model-specific, we selected several mainstream open-source large language models for evaluation, including Qwen-7B\cite{qwenQwen25TechnicalReport2025}, Mistral-7B\cite{jiangMistral7B2023}, and Llama-3.1-8B.

\textbf{Attack Method:} We adopt the PoisonedRAG\cite{zouPoisonedRAGKnowledgeCorruption2024} attack and prompt injection attack\cite{zhangHijackRAGHijackingAttacks2024} as our primary threat model to simulate adversarial conditions. This attack strategy aims to manipulate the retrieval system's output by injecting meticulously crafted poisoned documents into the knowledge base. 

\textbf{Baseline Methods:} To benchmark the performance of BiRD, we compare it against several representative and conceptually diverse defense methods from prior work:
\begin{itemize}
\item \textbf{Vanilla RAG:} The original, undefended RAG system serving as the vulnerable baseline.
\item \textbf{InstructRAG \cite{weiInstructRAGInstructingRetrievalAugmented2024}:} InstructRAG enhances the Retrieval-Augmented Generation process by employing self-synthesized rationales. It specifically instructs the language model to explain the derivation of the ground-truth answer from retrieved documents to improve internal reasoning. 
\item \textbf{RobustRAG \cite{xiangCertifiablyRobustRAG2024}:} This defense method is based on keyword extraction. Instead of providing the full set of retrieved documents as context, the defender extracts key phrases and provides only these keywords to the large language model. 
\item \textbf{ReliabilityRAG \cite{shenReliabilityRAGEffectiveProvably2025}:} This is a defense method based on solving the Maximum Independent Set (MIS) problem. The defender constructs a contradiction graph using the answers generated by the large language model, which are produced by pairing each document individually with the query. The maximum independent set is then solved within this graph, as this set corresponds to the subset of benign documents. 
\end{itemize}

\textbf{Implementation Details:} For the purpose of this experiment, we set the attacker's corruption size $m$ to 5. For our BiRD method, we set the filtering threshold $\epsilon$ to 2.5 and the number of initially retrieved documents $k$ to 20 in default.

\textbf{Metrics:} We employ Attack Success Rate (ASR) and Accuracy (ACC) as our primary evaluation metrics. For each response generated by the RAG system, the attack is considered successful if the output contains an exact string match to the adversary's target incorrect answer. Conversely, the response is considered correct if it matches the ground-truth answer. These metrics allow us to quantitatively assess both the vulnerability of the system under attack and the effectiveness of the defense methods in preserving answer quality.

\subsection{Baseline Experiments}
\label{subsec:baseline-experiments}
To systematically evaluate our approach, we conducted large-scale and comprehensive experiments. We selected VanillaRAG, InstructRAG, RobustRAG, and ReliabilityRAG as baselines and employed two attack methods, PoisonedRAG and PIA. For each configuration, we performed on three datasets, three retrievers, and three LLMs. The responses generated by the RAG system were evaluated via string matching against both the ground-truth answers and the target incorrect answers. 

As shown in Tables~\ref{tab:nq}, \ref{tab:msmarco}, and \ref{tab:hotpotqa}, BiRD achieves state-of-the-art performance across numerous baselines. Under the PoisonedRAG attack, BiRD reduces the ASR by up to 87\% and improves Acc by 55\% compared to VanillaRAG on the Mistral-NQ-ANCE configuration. In the face of PIA, although BiRD does not reach the absolute lowest ASR on Mistral-7B, it ensures a more practical response accuracy compared to ReliabilityRAG. Consequently, our approach achieves the optimal trade-off between defense robustness and system utility.

\subsection{Abalation Study}
To verify the necessity of the bidirectional retrieval score, we conducted ablation studies across multiple datasets. The experimental design evaluates performance metrics under three configurations: content-relevance-only filtering ($r_{cr}^i$, threshold 0.85), context-consistency-only filtering ($r_{cc}^i$, threshold 0.85), and the proposed composite score $S(d_i^q)$ (threshold 2.5).

Table~\ref{tab:ablation-bidirectional} shows a significant performance complementarity between the composite score and the single-signal variants. While content relevance ($r_{cr}^i$) maintains a marginal lead in Acc on HotpotQA and MSMARCO, reflecting its efficiency in capturing direct semantic relevance, its security remains highly vulnerable. In contrast, incorporating context consistency ($r_{cc}^i$) as a structural signal acts as a security filter, reducing the ASR on the NQ dataset by 0.37 compared to the $r_{cr}^i$-only strategy. The results indicate that the bidirectional framework achieves a substantial improvement in defensibility at the cost of minimal utility loss (<5\%), striking an optimal balance between security and practicality.
\label{subsec:ablation-study}
\begin{table}[!t]
\centering
\caption[Ablation study of bidirectional score]{Experimental results for bidirectional score abla-
tion. Best results ($\downarrow$ for ASR, $\uparrow$ for ACC) are in bold; second-best results are underlined.}
\label{tab:ablation-bidirectional}
\fontsize{8}{10}\selectfont
\setlength{\tabcolsep}{2.6pt}
\begin{tabularx}{\columnwidth}{@{} l c c *{3}{>{\centering\arraybackslash}X} @{} }
\toprule
\textbf{Setting} & $\mathbf{r_{cr}^i}$ & $\mathbf{r_{cc}^i}$ & \textbf{HotpotQA} & \textbf{MSMARCO} & \textbf{NQ} \\
\midrule
Forward Retrieval  & $\checkmark$ & {-} & \underline{0.34} / \textbf{0.70} & \underline{0.41} / \textbf{0.75} & 0.53 / \underline{0.68} \\
Backward Retrieval & {-} & $\checkmark$ & 0.52 / 0.62 & 0.42 / \underline{0.71} & \underline{0.44} / \textbf{0.70} \\
Bidirectional Retrieval & $\checkmark$ & $\checkmark$ & \textbf{0.24} / \underline{0.69} & \textbf{0.24} / \underline{0.71} & \textbf{0.16} / 0.62 \\
\bottomrule
\end{tabularx}
\end{table}
\begin{table}[!t]
\centering
\caption[Ranking similarity methods]{Experimental results for different rank similarity metrics. Best results ($\downarrow$ for ASR, $\uparrow$ for ACC) are in bold; second-best results are underlined.}
\label{tab:ranking-similarity-methods}
\fontsize{8}{10}\selectfont
\setlength{\tabcolsep}{2.6pt}
\begin{tabularx}{\columnwidth}{@{} l *{3}{>{\centering\arraybackslash}X} @{} }
\toprule
\textbf{Method} & \textbf{Jaccard} & \textbf{RBO} & \textbf{Spearman(Ours)} \\
\midrule
VanillaRAG & 0.93 / 0.29 & 0.93 / 0.29 & 0.93 / 0.29 \\
BiRD & \underline{0.37} / \underline{0.54} & 0.88 / 0.40 & \textbf{0.24} / \textbf{0.69} \\
$|\Delta|$ (BiRD $-$ Vanilla) & \underline{0.56}  / \underline{0.25}  & 0.05 / 0.11 & \textbf{0.69} / \textbf{0.40} \\
\bottomrule
\end{tabularx}
\end{table}
\subsection{Dynamic Performance Experiments}
\label{subsec:dynamic-performance-experiments}
\noindent\textbf{Impact of context-consistency metric $\varrho(\cdot,\cdot)$.} To evaluate the influence of different measurement methods of the similarity between two rankings (used to compute $r_{cc}^i$) on the performance of our BiRD defense mechanism, we conducted a series of experiments using three widely-adopted metrics: Jaccard distance\cite{https://doi.org/10.1111/j.1469-8137.1912.tb05611.x}, Spearman's rank correlation coefficient\cite{fiellerTESTSRANKCORRELATION1957}, and Rank-biased Overlap (RBO)\cite{webberSimilarityMeasureIndefinite2010a}.

As shown in Table~\ref{tab:ranking-similarity-methods}, the BiRD defense mechanism significantly reduces both the ASR and ACC when Spearman's rank correlation coefficient is employed as the ranking similarity metric. In contrast, when using Jaccard distance or RBO, the defense fails to effectively decrease the ASR and ACC. Furthermore, the ASR actually increases when the RBO method is applied.

\noindent\textbf{Impact of filtering threshold $\epsilon$.}
\begin{figure*}[ht]
\centering 
\includegraphics[width=\textwidth]{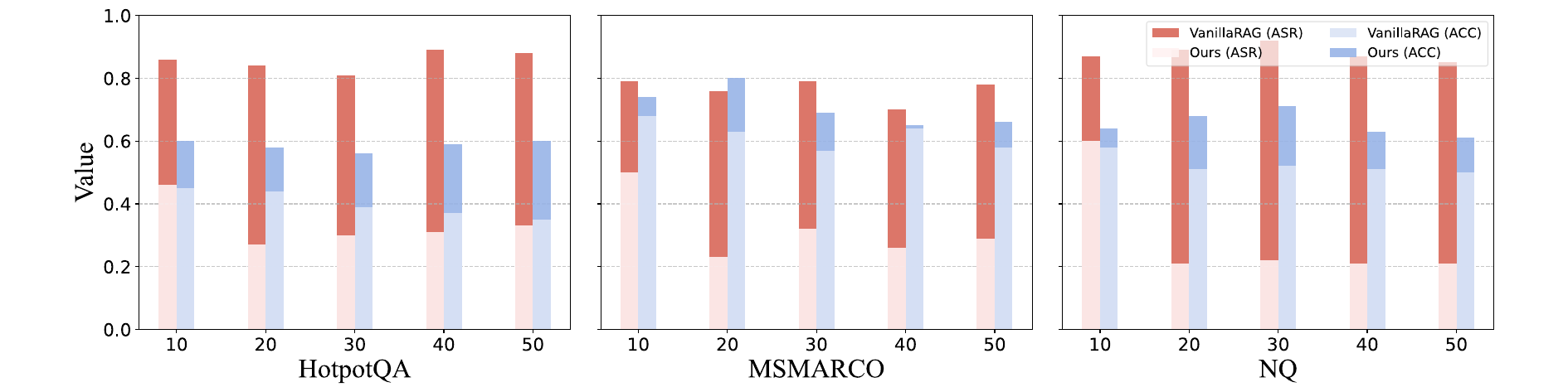}
\caption{The variation of ASR and ACC across different datasets in relation to the parameter $k$ in BiRD, where the difference between the bars at the same position in the clustered bar chart represents the magnitude of our reduction or improvement.}
\label{fig:k_results}
\end{figure*}
To evaluate the sensitivity and operational characteristics of the BiRD defense mechanism regarding the filtering threshold, we conducted a series of experiments by varying $\epsilon$ within a reasonable range of $[1, 3.5]$.

\begin{figure}[t]
\centering
\includegraphics[width=\columnwidth]{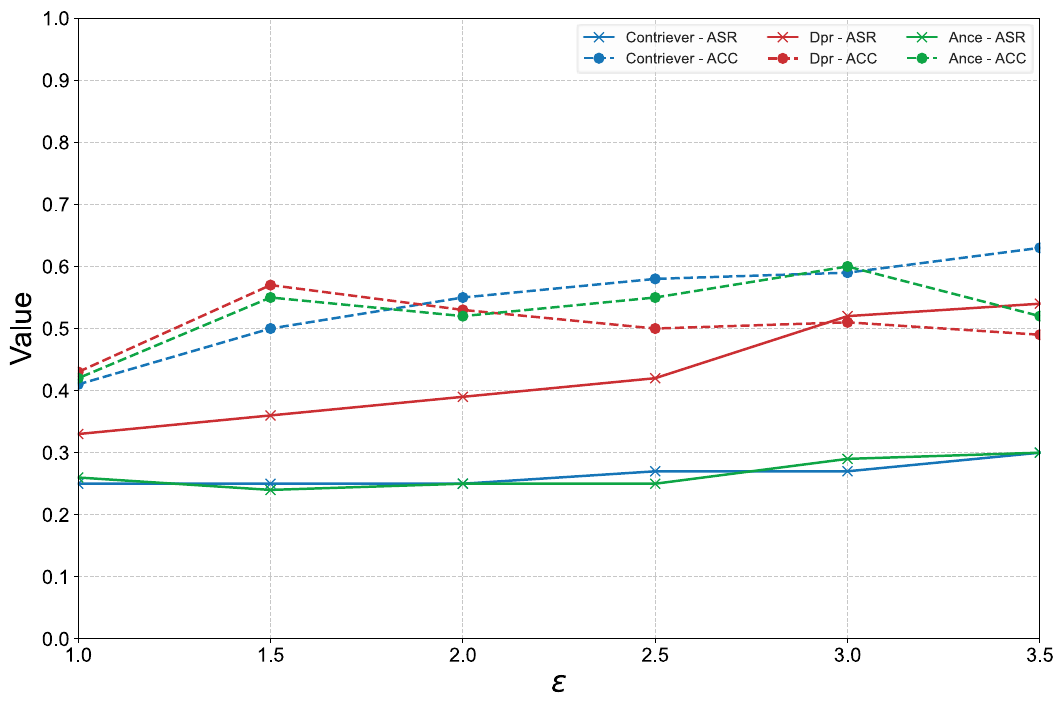}
\caption{ASR and ACC versus the filtering threshold $\epsilon$ in BiRD.}
\label{fig:threshold_results}
\end{figure}

As illustrated in Figure \ref{fig:threshold_results}, the results reveal a clear and compelling trade-off curve. Through a holistic evaluation of both metrics, we determined that the interval between $\epsilon = 2$ and $\epsilon = 3$ offers the most favorable balance. Within this range, the system maintains a high accuracy (ACC $>$ 0.6) while keeping the attack success rate at a relatively controlled level (ASR $<$ 0.25).

In practical applications, the selection of an appropriate threshold must be based on the specific trade-off between security and utility. A lower $\epsilon$ signifies a more stringent defense that may discard a greater number of documents, including clean ones. Conversely, a higher $\epsilon$ represents a more permissive defense strategy that may include poisoned documents but offers higher system utility.

\noindent\textbf{Impact of retrieval size $k$.}
\begin{figure}[t]
\centering
\includegraphics[width=\columnwidth]{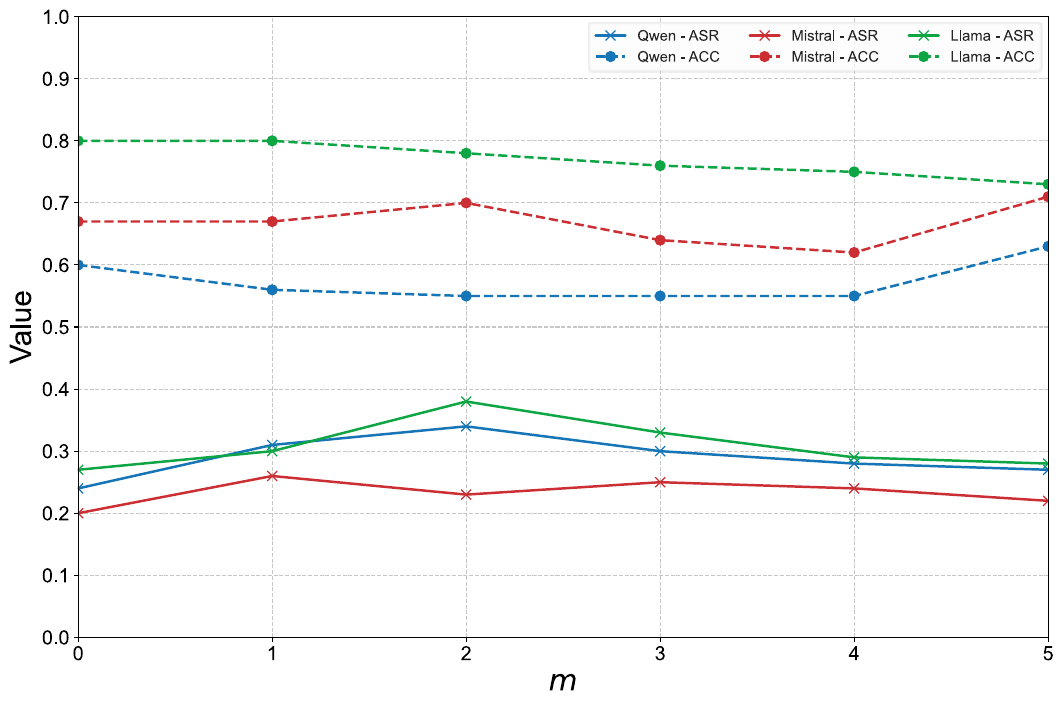}
\caption{The variation of ASR and ACC across different datasets in relation to the parameter $m$ for attack.}
\label{fig:m_results}
\end{figure}
To evaluate the impact of the retrieval size $k$ on the robustness of our defense method, we conducted a series of experiments on the HotpotQA, MSMARCO, and NQ datasets. We employed Qwen and Contriever as the LLM and retriever respectively while varying the parameter $k$ within a reasonable range from 10 to 50.

As illustrated in Figure \ref{fig:k_results}, BiRD consistently reduces the ASR and improves the ACC to varying degrees across all tested values of $k$. Our approach remains effective in lowering the ASR even when $k$ is large. This performance demonstrates that BiRD can successfully filter out poisoned documents even in scenarios where they occupy a very small proportion of the total results, such as when only $m=5$ poisoned documents are present within an initial retrieval set of $k=50$.

\noindent\textbf{Impact of corruption size $m$.} To evaluate the impact of the corruption size $m$, which represents the number of poisoned documents targeting a single query, on the robustness of our defense method, we conducted a series of experiments using three large language models: Qwen, Mistral, and Llama. These experiments utilized HotpotQA as the dataset and Contriever as the retriever while varying the parameter $m$ within a reasonable range from 0 to 5.

As illustrated in Figure \ref{fig:m_results}, BiRD maintains high robustness against variations in the corruption size. Even as $m$ increases, our defense mechanism keeps the performance of the RAG system within a secure and reliable range. In contrast, VanillaRAG exhibits a significant increase in ASR and a notable decline in ACC even when faced with only a single poisoned document. These results further highlight the inherent vulnerability of standard RAG systems and demonstrate the effectiveness of our proposed defense.

\begin{table}[ht]
\centering
\caption{Efficiency Analysis: Average Runtime (s/query) on three datasets. Best results ($\downarrow$ for ASR, $\uparrow$ for ACC) are in bold; second-best results are underlined.}
\label{tab:runtime-overhead}
\fontsize{7}{8.5}\selectfont
\setlength{\tabcolsep}{0.4pt}
\newcolumntype{C}{>{\centering\arraybackslash}X}

\begin{tabularx}{\columnwidth}{@{} l *{9}{C} @{}}
\toprule
\multirow{2}{*}{\textbf{Method}} & \multicolumn{3}{c}{\textbf{NQ}} & \multicolumn{3}{c}{\textbf{HotpotQA}} & \multicolumn{3}{c}{\textbf{MSMARCO}} \\
\cmidrule(lr){2-4} \cmidrule(lr){5-7} \cmidrule(lr){8-10}
& Cont. & ANCE & DPR & Cont. & ANCE & DPR & Cont. & ANCE & DPR \\
\midrule

\rowcolor[HTML]{F2F2F2} \multicolumn{10}{c}{\textbf{Model: Qwen2.5-7B-Instruct}} \\
InstructRAG & \underline{5.57} & \underline{6.07} & \underline{5.69} & \underline{5.65} & \underline{6.18} & \underline{5.80} & \underline{5.36} & \underline{5.67} & \underline{5.71} \\
ReliabilityRAG & 8.92 & 28.30 & 27.98 & 25.90 & 26.53 & 26.61 & 25.72 & 25.53 & 25.48 \\
RobustRAG & 25.73 & 23.32 & 23.24 & 19.73 & 23.43 & 21.81 & 31.72 & 22.09 & 21.68 \\
\textbf{BiRD (Ours)} & \textbf{3.00} & \textbf{3.61} & \textbf{3.00} & \textbf{3.00} & \textbf{6.05} & \textbf{5.44} & \textbf{3.00} & \textbf{4.83} & \textbf{4.22} \\
\midrule

\rowcolor[HTML]{F2F2F2} \multicolumn{10}{c}{\textbf{Model: Mistral-7B}} \\
InstructRAG & \underline{4.93} & \textbf{4.55} & \textbf{4.82} & \textbf{4.67} & \textbf{3.98} & \textbf{3.86} & \textbf{4.79} & \underline{4.92} & \textbf{4.34} \\
ReliabilityRAG & 13.37 & 14.31 & 13.80 & 14.68 & 16.62 & 14.61 & 11.18 & 11.89 & 11.23 \\
RobustRAG & 10.21 & 12.51 & 11.02 & 12.56 & 12.87 & 12.24 & 8.50 & \underline{7.83} & \underline{8.46} \\
\textbf{BiRD (Ours)} & \textbf{3.61} & \underline{6.05} & \underline{4.83} & \underline{4.93} & \underline{7.37} & \underline{6.15} & \underline{7.27} & \textbf{3.71} & 8.49 \\
\midrule

\rowcolor[HTML]{F2F2F2} \multicolumn{10}{c}{\textbf{Model: Llama-3.1-8B}} \\
InstructRAG & \underline{4.91} & \textbf{4.96} & \underline{5.54} & \underline{5.54} & \textbf{5.23} & \textbf{5.21} & \textbf{6.22} & \underline{6.13} & \textbf{5.73} \\
ReliabilityRAG & 39.49 & 59.88 & 40.79 & 38.01 & 40.01 & 38.69 & 38.15 & 38.05 & 37.65 \\
RobustRAG & 46.59 & 46.55 & 44.98 & 23.56 & 48.45 & 23.15 & 21.30 & 22.48 & 21.91 \\
\textbf{BiRD (Ours)} & \textbf{3.00} & \underline{5.44} & \textbf{4.22} & \textbf{4.32} & \underline{6.76} & \underline{5.54} & \underline{6.66} & \textbf{6.05} & \underline{7.88} \\
\bottomrule
\end{tabularx}
\end{table}
\subsection{Runtime Experiments}
\label{subsec:runtime-experiments}
In addition to evaluating defense effectiveness and utility preservation, we conducted a comprehensive runtime analysis to assess the practical efficiency overhead introduced by different defense mechanisms. In the real-world deployment of secure RAG systems, computational cost is a critical factor.

Using the same hardware configuration of four A6000 48GB GPUs, we measured the average end-to-end inference time per query for each defense method across three evaluated LLMs, three datasets, and three retrievers. The reported time encompasses document retrieval, defense processing, and the LLM generation process.

The runtime comparison is presented in Table \ref{tab:runtime-overhead}. The results reveal significant efficiency differences among the various defense strategies. Compared to previous work, our proposed BiRD method introduces only a minimal overhead. On average, it adds a mere 0.9s of additional latency, which stands in stark contrast to the extra overhead of tens of seconds required by other methods. This demonstrates that the bidirectional ranking and filtering process of BiRD is highly efficient.

The runtime experiments confirm that BiRD achieves an ideal balance between security and efficiency. While providing robust defense, it maintains a latency profile very close to the vulnerable Vanilla RAG baseline and is significantly more efficient than alternative defense methods.
\label{subsec:runtime_experiments}

\subsection{Validation-Set Threshold Selection}
\label{subsec:validation-threshold-selection}
\textbf{Goal and setup.} To validate the practicality of the threshold selection strategy described in Section~4.2.3, we construct a validation set on MSMARCO using poisoned documents from PoisonedRAG. Specifically, we treat PoisonedRAG-poisoned documents as the validation set, compute the score $S(d_i^q)$ for poisoned documents, and estimate their score distribution. We then select a fixed threshold $\epsilon=2.5$ according to this distribution. Importantly, we do \emph{not} use any PIA-poisoned documents in this calibration process. After selecting $\epsilon$, we directly apply it to defend against PIA on MSMARCO under the Qwen2.5-7B model and three retrievers. The defense results are summarized in Table~\ref{tab:validation-threshold-msmarco}.

\textbf{Analysis.} As shown in Table~\ref{tab:validation-threshold-msmarco}, the threshold obtained from PoisonedRAG validation generalizes well to the unseen PIA setting. Across all three retrievers, BiRD achieves the lowest ASR and the highest ACC among the compared defenses, indicating that calibrating $\epsilon$ using PoisonedRAG-poisoned documents is a feasible and effective deployment strategy. This also suggests that the score distribution of poisoned documents provides a stable signal for selecting $\epsilon$ without requiring access to the specific attack type encountered at test time.

\begin{table}[t]
\centering
\caption{PIA defense on MSMARCO with $\epsilon=2.5$ selected from PoisonedRAG validation-set. Best results ($\downarrow$ for ASR, $\uparrow$ for ACC) are in bold; second-best results are underlined.}
\label{tab:validation-threshold-msmarco}
\fontsize{8}{10}\selectfont
\setlength{\tabcolsep}{2.5pt}
\begin{tabularx}{\columnwidth}{@{} l *{3}{>{\centering\arraybackslash}X} @{} }
\toprule
\textbf{Baseline} & \textbf{Contriever} & \textbf{ANCE} & \textbf{DPR} \\
\midrule
RobustRAG & 0.71 / 0.61 & 0.69 / 0.52 & 0.65 / 0.58 \\
InstructRAG & 0.53 / 0.55 & 0.63 / 0.56 & 0.50 / 0.55 \\
ReliabilityRAG & \underline{0.30} / \underline{0.62} & \underline{0.27} / \underline{0.59} & \underline{0.31} / \underline{0.59} \\
BiRD (Ours) & \textbf{0.27} / \textbf{0.63} & \textbf{0.22} / \textbf{0.69} & \textbf{0.25} / \textbf{0.63} \\
\bottomrule
\end{tabularx}
\end{table}


\section{Conclusions}

In conclusion, our proposed BiRD defense method introduce and utilizes context consistency. The proposed defense strategy is highly efficient, with minimal overhead compared to other defense methods.

For future work, we plan to explore adaptive thresholding techniques to further enhance the balance between security and utility. Additionally, we aim to investigate the applicability of our defense mechanism in other retrieval-augmented tasks beyond question answering, such as summarization and dialogue systems.
\section{Acknowledgments}
We thank the anonymous reviewers for their constructive feedback. We also thank the maintainers of the open-source retrievers and LLMs used in our evaluation, as well as the authors who released the datasets and baselines that enabled reproducible comparisons.

\section*{Ethical Considerations}
We have reviewed the USENIX Security Ethics Guidelines. This work aligns with its principles as it presents a defensive security technique and aims to improve the robustness of RAG systems against corpus poisoning.

\noindent\textbf{Human subjects and sensitive data.} Our study does not involve human subjects experiments and does not collect personal or sensitive user data. All experiments are performed on public datasets and pre-trained models.

\noindent\textbf{Dual-use and misuse risk.} While our analysis discusses attack behaviors for evaluation purposes, we focus on mitigation and do not provide operational guidance for deploying real-world attacks. We encourage responsible disclosure and careful deployment of RAG systems in safety-critical settings.

\noindent\textbf{Bias and fairness.} Corpus poisoning may amplify biases present in data sources. Our defense is designed to reduce the impact of maliciously injected content, but it is not a complete solution to data bias; we recommend complementary bias auditing when deploying RAG.

\section*{Open Science}
We commit to releasing the artifacts necessary to reproduce our results, including (1) the implementation of BiRD, (2) scripts for running attacks and defenses, and (3) evaluation code for ASR/ACC and runtime measurement.

The datasets are available at: \url{https://public.ukp.informatik.tu-darmstadt.de/thakur/BEIR/datasets/}, 
The models are available at: \url{https://huggingface.co/}.

\bibliographystyle{IEEEtran}
\bibliography{refs}
\appendix
\section{Appendix}
\subsection{Formulation of Bidirectional Ranking Defense}
To facilitate a clear understanding of the proposed framework, we provide a comprehensive summary of the mathematical notations and variables used throughout this paper. Table~\ref{tab:notations} categorizes the symbols related to the poisoning attack scenario, the RAG process, and the specific parameters of BiRD method. This serves as a quick reference for the formal definitions of the attack surface and the defensive scoring mechanisms.
\begin{table}[!t]
\centering
\caption{Summary of Variables and Mathematical Notations}
\label{tab:notations}
\begin{tabularx}{\columnwidth}{@{} l X @{}}
\toprule
\textbf{Symbol} & \textbf{Description} \\
\midrule
$\mathcal{D}_{\text{benign}}$ & Original benign document corpus \\
$\mathcal{D}_{\text{poison}}$ & Injected poisoned document set \\
$\mathcal{D}'$ & Poisoned corpus $\mathcal{D}_{\text{benign}} \cup \mathcal{D}_{\text{poison}}$ \\
$N$ & Total number of documents in $\mathcal{D}'$ \\
$k$ & Number of top documents retrieved by the system \\
$R(\cdot,\cdot)$ & Retriever function mapping (query, corpus) to a ranked list \\
$\text{top-}k(\cdot)$ & Function that returns the top-$k$ ranked documents \\
$d_i^q$ & The $i$-th doc. in the forward ranking of query $q$ \\
$d_j^i$ & The $j$-th doc. in the backward ranking of query $d_i^q$ \\
$\text{LLM}(\cdot,\cdot)$ & Large Language Model function \\
$\mathcal{D}_{\text{clean}}$ & Filtered document set provided to the LLM as context \\
$R_{\text{fw}}^q$ & Forward ranking list for query $q$ \\
$R_{\text{bw}}^i$ & Backward ranking list using $d_i^q$ as a query \\
$r_{cr}^i$ & Content relevance of $d_i^q$ (forward similarity) \\
$r_{cc}^i$ & Context consistency of $d_i^q$ (ranking similarity) \\
$\text{Sim}(\cdot,\cdot)$ & Similarity function used by the retriever \\
$\varrho(\cdot,\cdot)$ & Spearman's rank correlation coefficient \\
$\mathcal{C}$ & Common documents between $R_{\text{fw}}^q$ and $R_{\text{bw}}^i$ \\
$r_f(d), r_b(d)$ & Rank of $d$ in forward and backward rankings \\
$S(d_i^q)$ & Composite score for $d_i^q$ \\
$\epsilon$ & Defense threshold for filtering \\
\bottomrule
\end{tabularx}
\end{table}

\subsection{Preliminary Experiments} 
\label{app:preliminary_experiments}
\begin{figure*}[t]
    \centering
    \includegraphics[width=0.8\textwidth]{./figs/average_legend.pdf} \\
    \vspace{2mm}
    \subfloat[HotpotQA-ANCE]{
        \includegraphics[width=0.31\textwidth]{./figs/hotpotqa_ance_results_heatmap.pdf}
    } \hfill
    \subfloat[HotpotQA-Contriever]{
        \includegraphics[width=0.31\textwidth]{./figs/hotpotqa_contriever_results_heatmap.pdf}
    } \hfill
    \subfloat[HotpotQA-DPR]{
        \includegraphics[width=0.31\textwidth]{./figs/hotpotqa_dpr_results_heatmap.pdf}
    }

    \subfloat[MSMARCO-ANCE]{
        \includegraphics[width=0.31\textwidth]{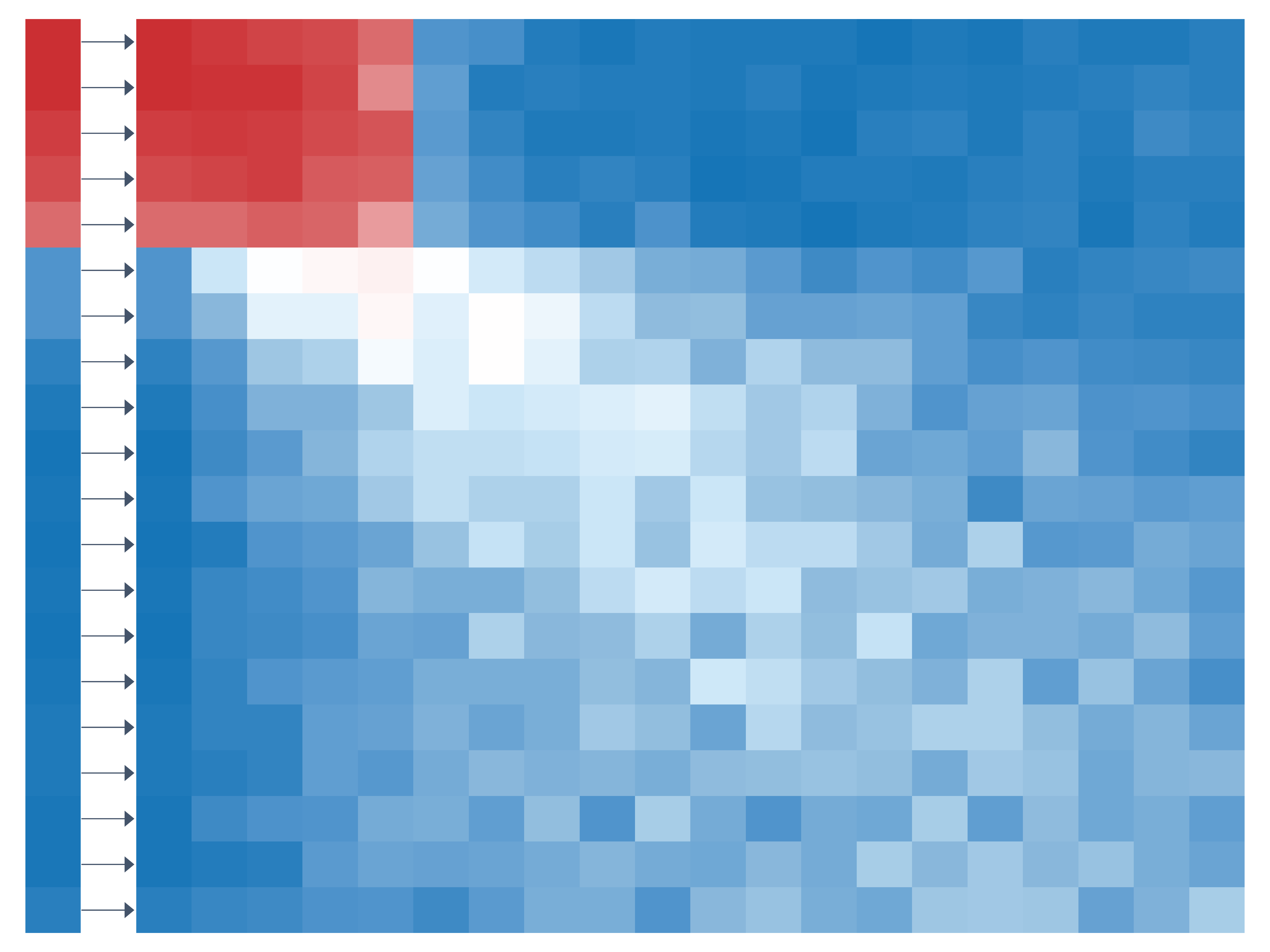}
    } \hfill
    \subfloat[MSMARCO-Contriever]{
        \includegraphics[width=0.31\textwidth]{./figs/msmarco_contriever_results_heatmap.pdf}
    } \hfill
    \subfloat[MSMARCO-DPR]{
        \includegraphics[width=0.31\textwidth]{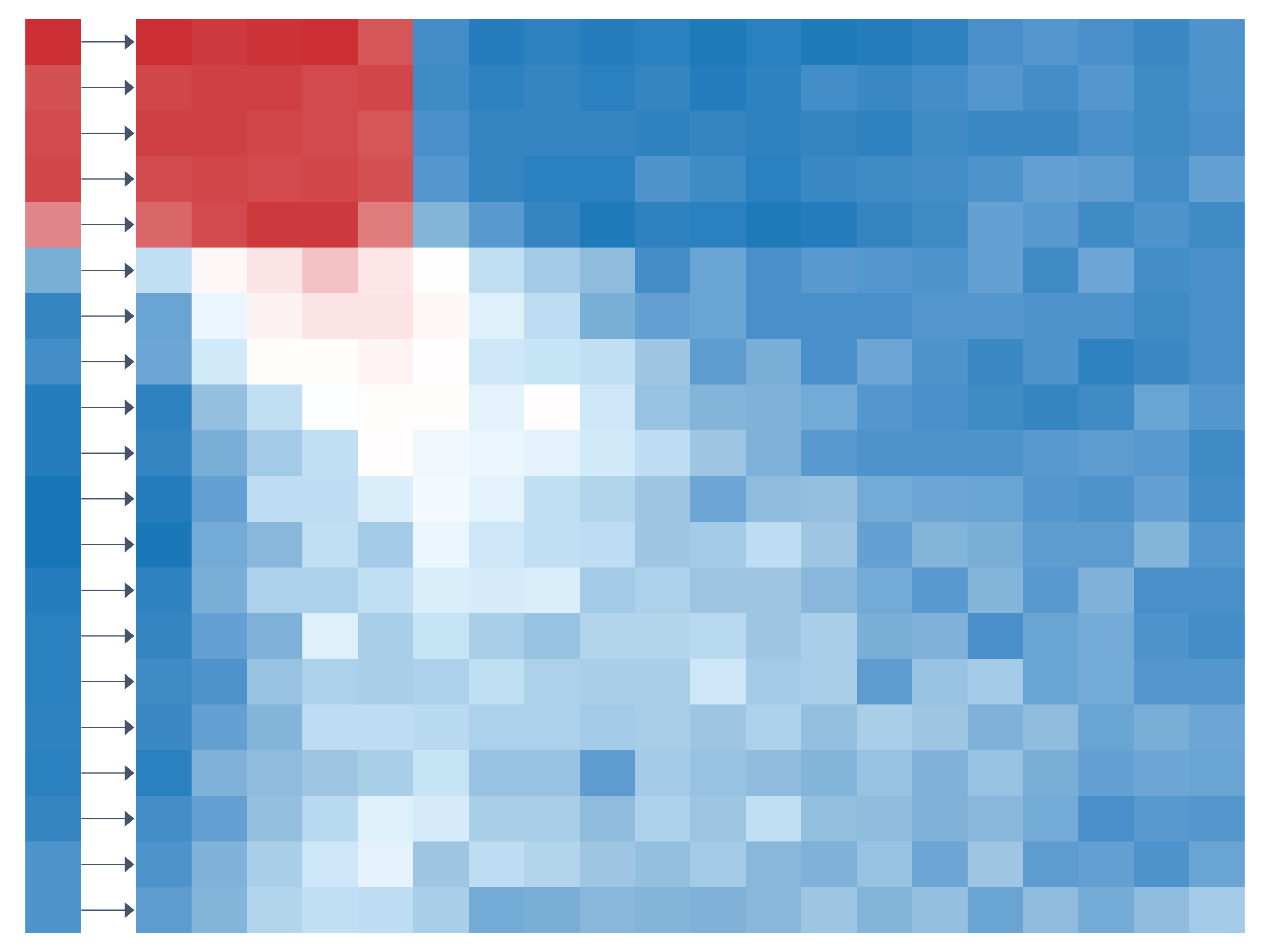}
    }

    \subfloat[NQ-ANCE]{
        \includegraphics[width=0.31\textwidth]{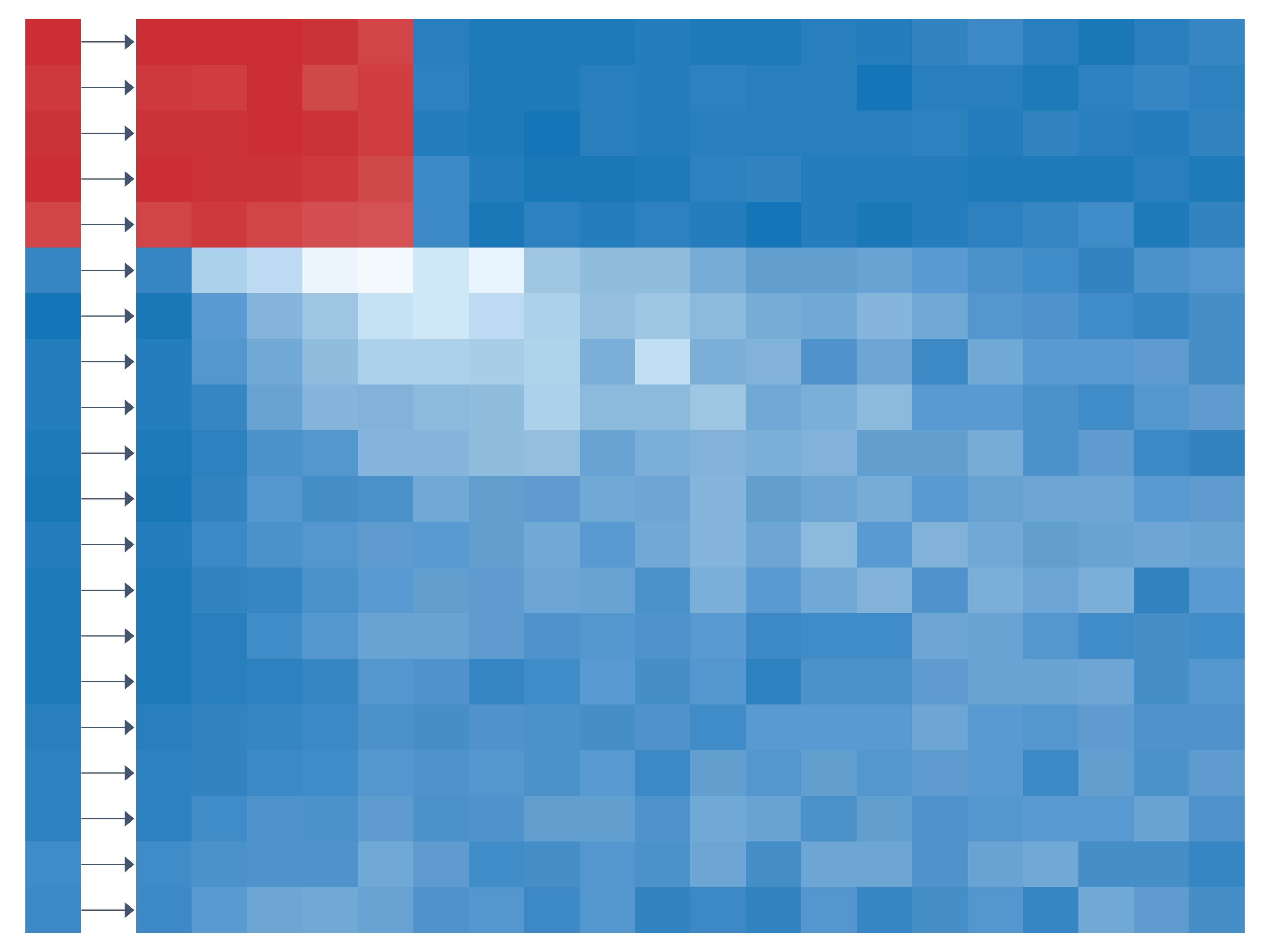}
    } \hfill
    \subfloat[NQ-Contriever]{
        \includegraphics[width=0.31\textwidth]{./figs/nq_contriever_results_heatmap.pdf}
    } \hfill
    \subfloat[NQ-DPR]{
        \includegraphics[width=0.31\textwidth]{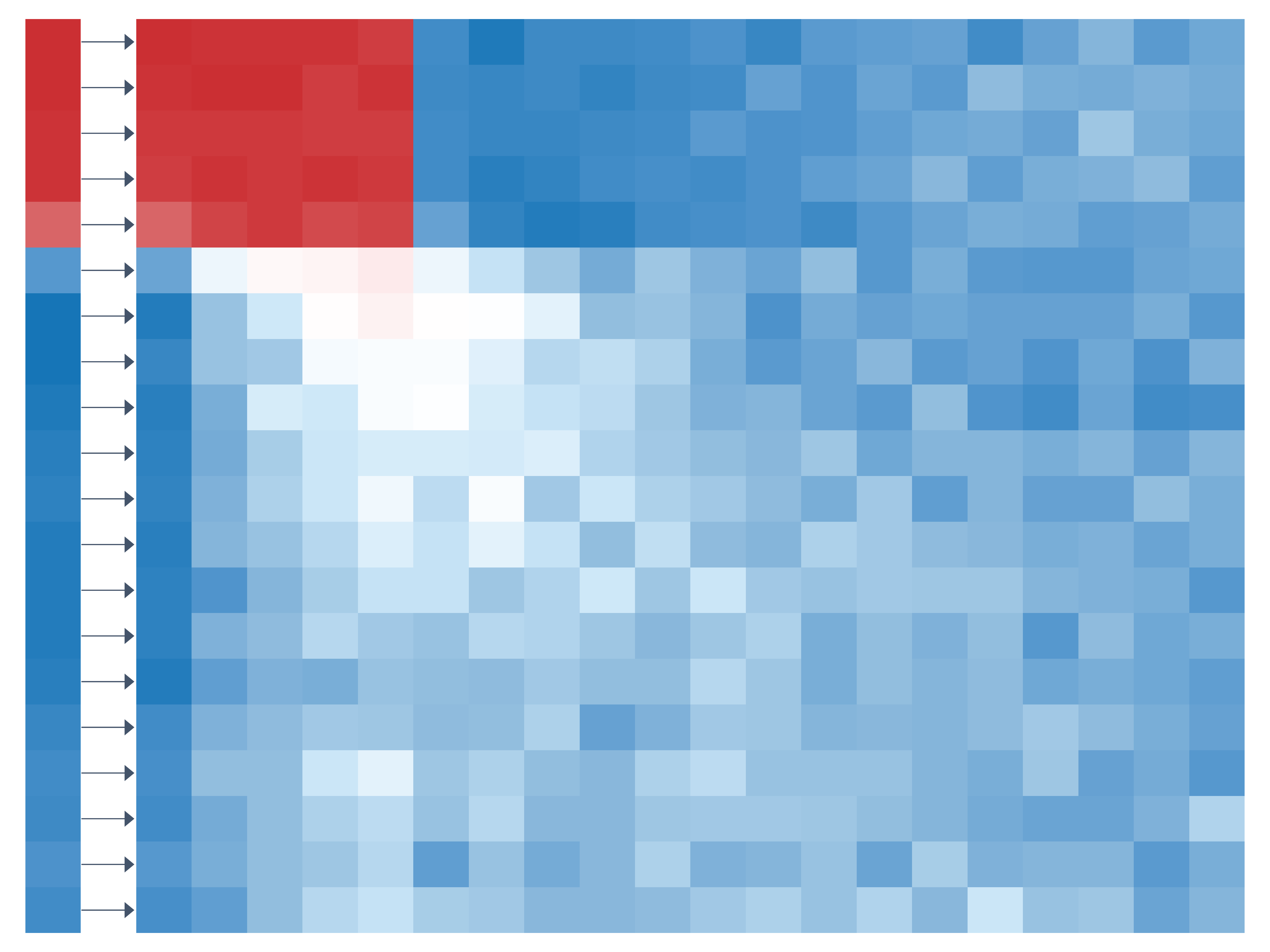}
    }

    \caption{Comprehensive comparison of rank position poisoned frequency heatmaps across three datasets (rows) and three retrievers (columns).}
    \label{fig:3x3_full_matrix}
\end{figure*}

\begin{figure*}[t]
    \centering
    \includegraphics[width=0.8\textwidth]{./figs/average_legend.pdf} \\
    \vspace{2mm}

    \subfloat[Topic-Flip-PRO (ANCE)]{
        \includegraphics[width=0.31\textwidth]{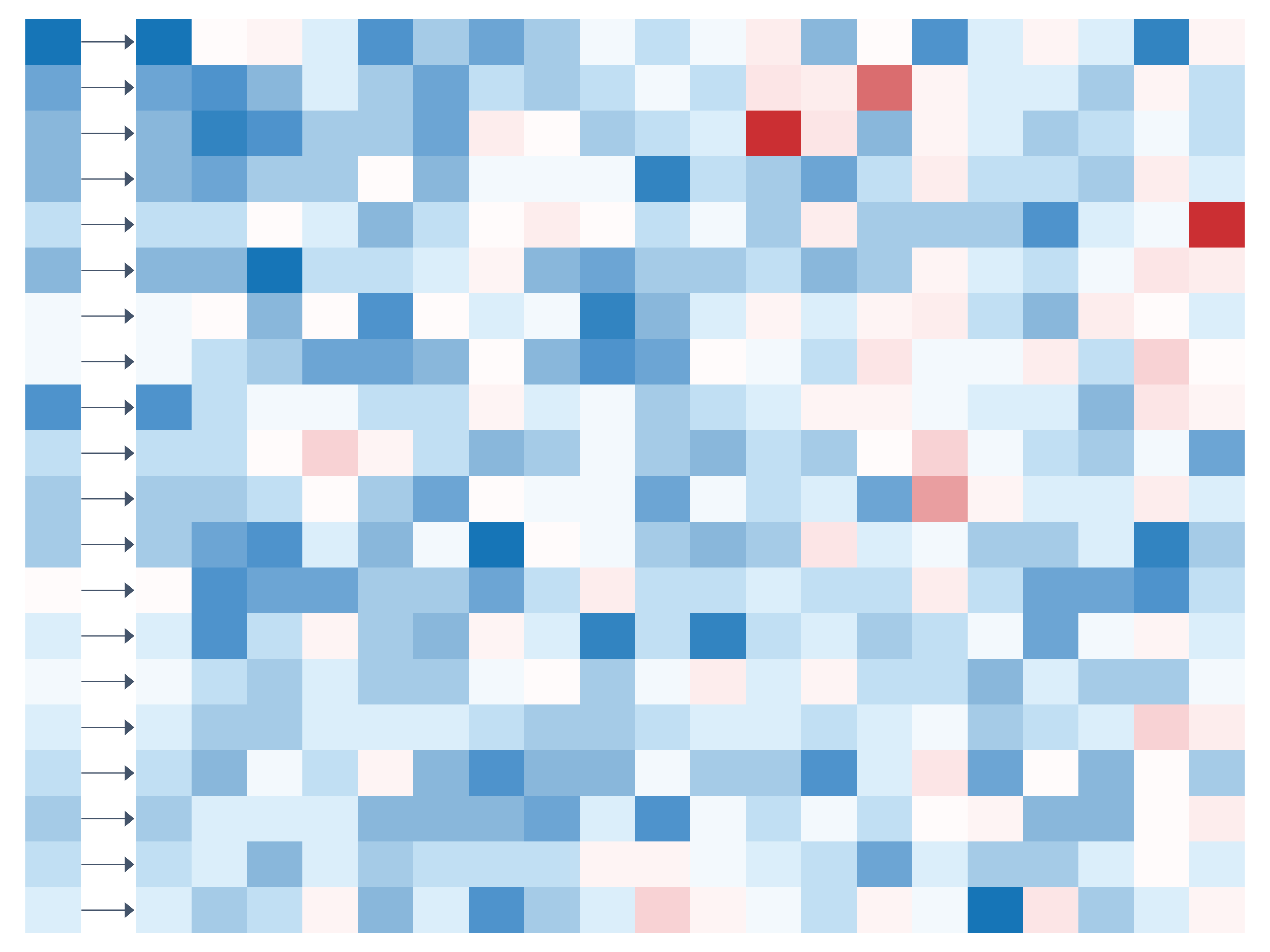}
    } \hfill
    \subfloat[Topic-Flip-PRO (Contriever)]{
        \includegraphics[width=0.31\textwidth]{./figs/Topic_PROCON_data_contriever_PRO_results_heatmap.pdf}
    } \hfill
    \subfloat[Topic-Flip-PRO (DPR)]{
        \includegraphics[width=0.31\textwidth]{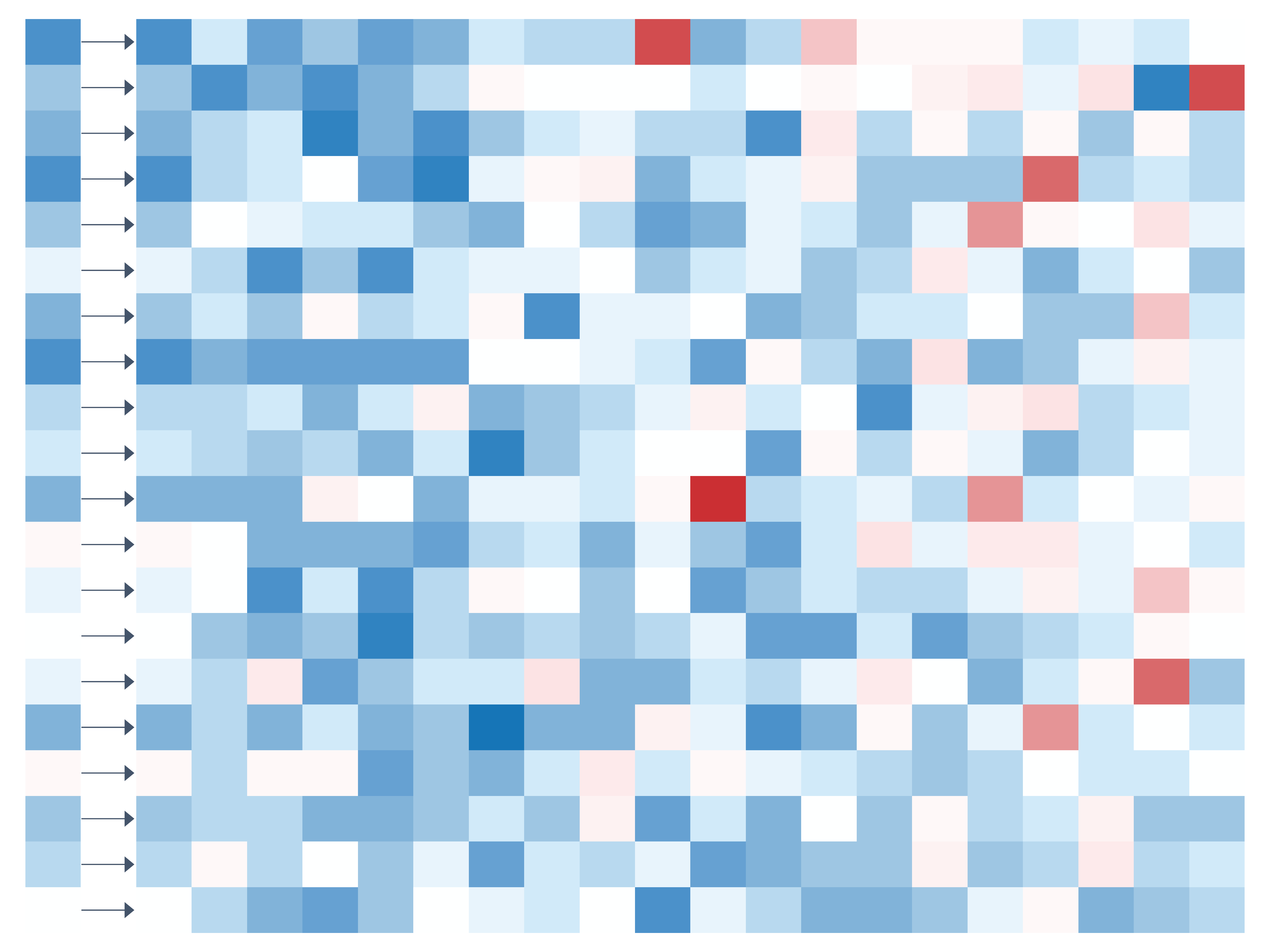}
    }

    \vspace{2mm} 

    \subfloat[Topic-Flip-CON (ANCE)]{
        \includegraphics[width=0.31\textwidth]{./figs/Topic_PROCON_data_ance_CON_results_heatmap.pdf}
    } \hfill
    \subfloat[Topic-Flip-CON (Contriever)]{
        \includegraphics[width=0.31\textwidth]{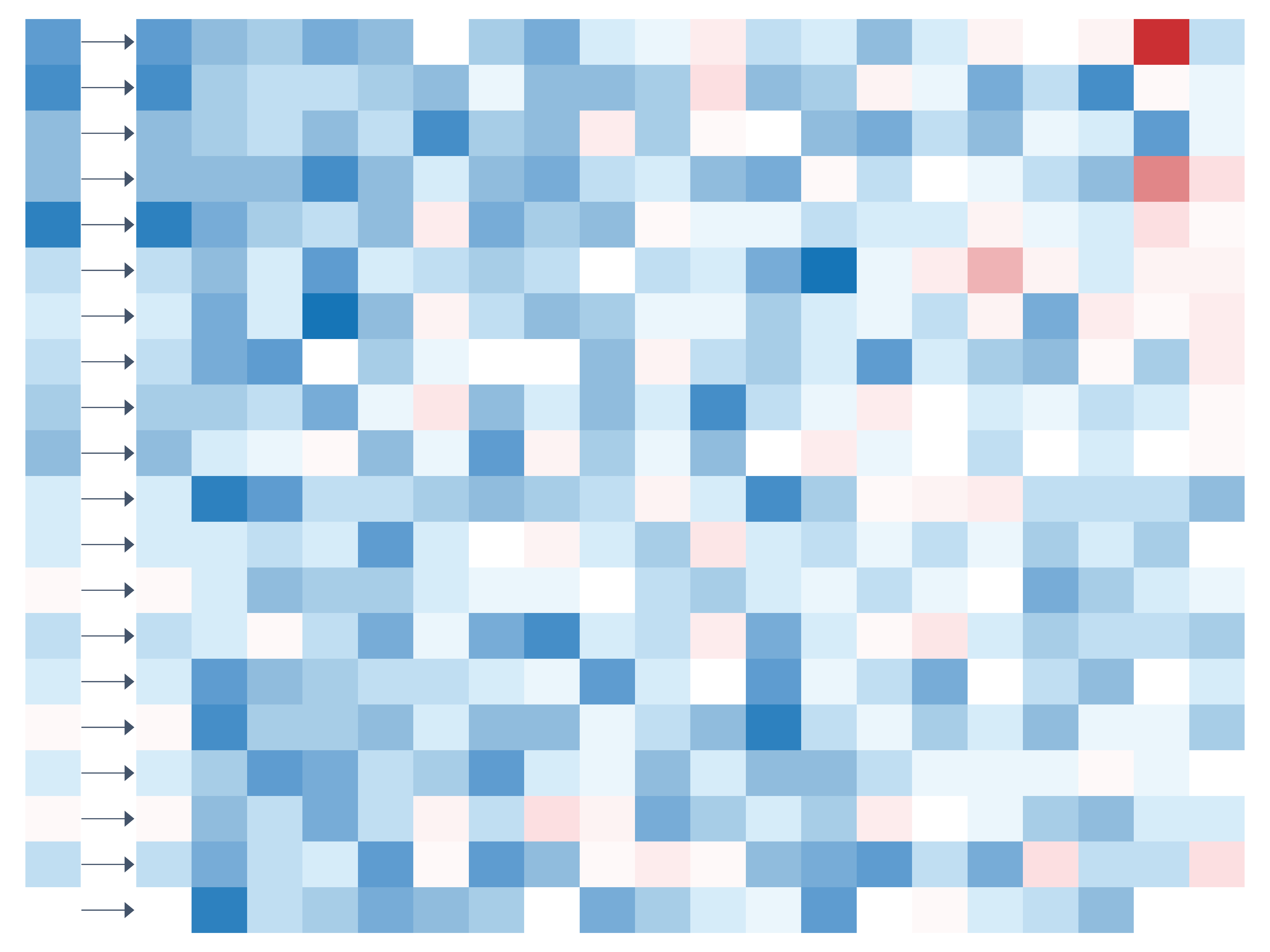}
    } \hfill
    \subfloat[Topic-Flip-CON (DPR)]{
        \includegraphics[width=0.31\textwidth]{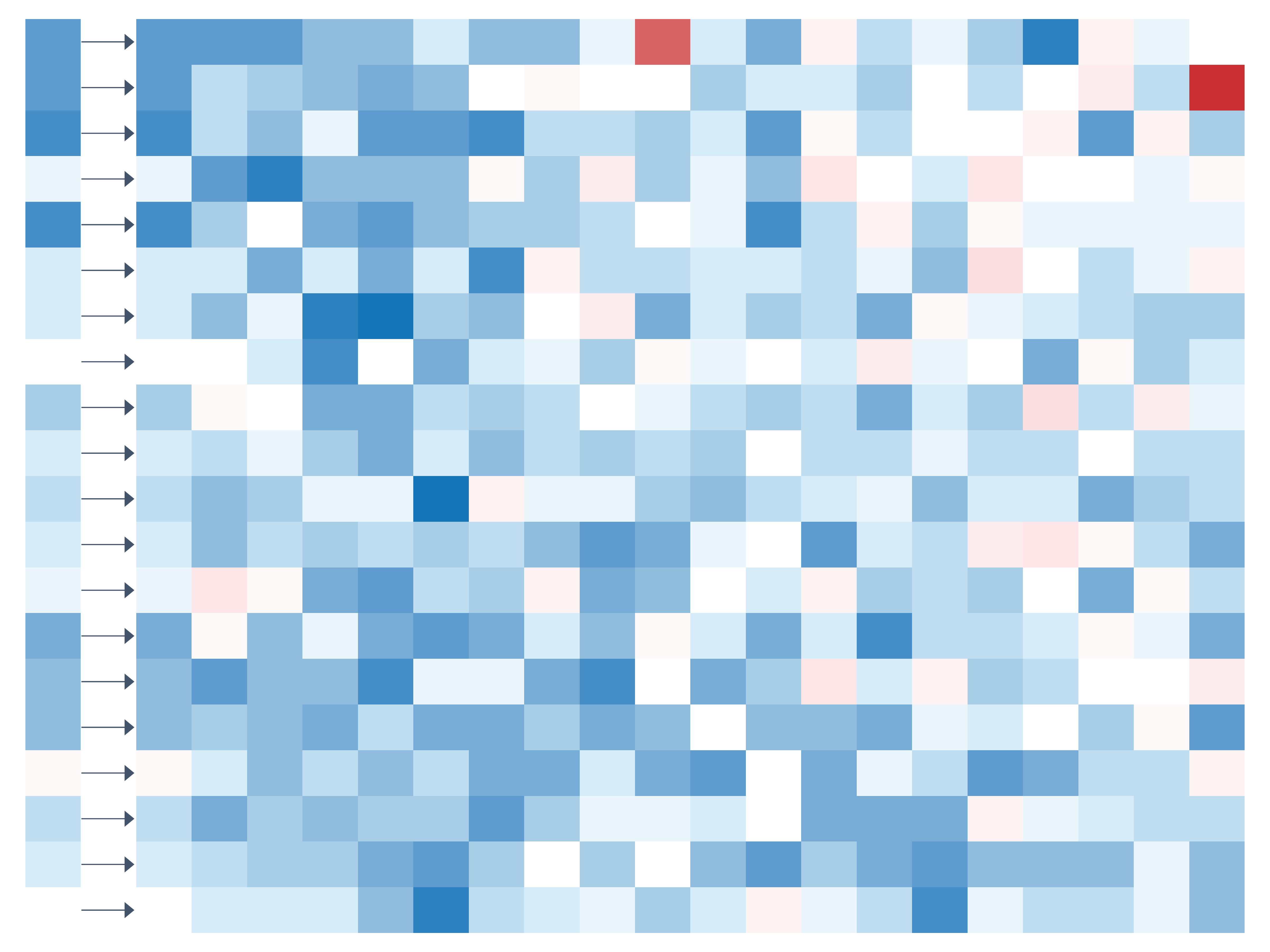}
    }

    \caption{Rank position poison frequency for Topic-Flip attacks: PRO strategy (top row) and CON strategy (bottom row) across ANCE, Contriever, and DPR retrievers.}
    \label{fig:pro_con_comparison}
\end{figure*}

We perform a series of preliminary experiments to evaluate the effectiveness of our approach. Figure \ref{fig:3x3_full_matrix} is a 3x3 matrix that shows the discovery under PoisonedRAG attack. Figure \ref{fig:pro_con_comparison} is a 2x3 matrix that shows the discovery under Topic-FlipRAG attack.
\end{document}